\pdfoutput=1

\documentclass[final,3p,times]{elsarticle}
\usepackage[latin9]{inputenc}
\usepackage{bm}
\usepackage{amsmath}
\usepackage{amssymb}
\usepackage{graphicx}
\usepackage{esint}
\usepackage{multirow}
\usepackage[subfigure]{graphfig}
\usepackage[figuresright]{rotating}
\usepackage{verbatim}
\usepackage{setspace}
\usepackage{lineno,hyperref}
\makeatletter

\usepackage{tabularx}\usepackage{subfigure}\usepackage{subfigure}\usepackage{color}

\usepackage{setspace}

\usepackage{bm}

\journal{Journal of Computational Physics}

\makeatother

\begin{document}
\title{Coupled THINC and level set method: a conservative interface capturing scheme with arbitrary-order surface representations}

\author[ad1,ad2]{Longgen Qian}

\author[ad1]{Yanhong Wei \corref{cor}}

\author[ad2]{Feng Xiao \corref{cor}}

\address[ad1]{College of Materials Science and Technology, Nanjing University of Aeronautics and Astronautics, Nanjing, 211100, China}

\address[ad2]{Department of Mechanical Engineering, Tokyo Institute of Technology,4259 Nagatsuta Midori-ku, Yokohama, 226-8502, Japan.}

\cortext[cor]{Corresponding author: Dr. Yanhong. Wei (Email: nuaadw@126.com); Dr. F. Xiao (Email: xiao.f.aa@m.titech.ac.jp)}.

\begin{abstract}
\indent In this paper, we propose a simple and accurate numerical method for capturing moving interfaces on fixed Eulerian grids by coupling the Tangent of Hyperbola Interface Capturing (THINC) method and Level Set (LS) method. The innovative and practically-significant aspects of the proposed method, so-called THINC/LS method, lie in (1) representing the interface with polynomial of arbitrary-order rather than the plane representation commonly used in the Piecewise Linear Interface Calculation (PLIC) Volume of fluid (VOF) methods, (2) conserving rigorously the mass of the transported VOF field, and (3) providing a straightforward and easy-to-code algorithm for implementation. We  verified the proposed scheme with the widely used benchmark tests. Numerical results show that this new method can obtain arbitrary-order accuracy for interface reconstruction and can produce more accurate results than the classical VOF methods. 
\end{abstract}

\maketitle

\section{Introduction}

Moving interface phenomena, such as multiphase flows, moving fronts with chemical reactions or phase change, are widely found in nature and engineering fields. Numerical modeling and computation of moving interfaces have a long history of researches with large amount efforts devoted, but still remain a challenging and unresolved topic. Numerous methods have been proposed to identify and compute moving interfaces by using either marking points \cite{Tryggvason1992,Tryggvason2001,Rider1995a,Rider1995b} or identification functions. The latter approach solves the identification function for moving interface on fixed Eulerian grids, and is efficient and robust for interface which experiences even topological changes, like merging and breaking. The most popularly used identification functions are the volume fraction or volume of fluid (VOF) function   \cite{Hirt1981,Youngs1982,Lafaurie1994,Rudman1997,Rider1998,Puckett2004} and the level set (LS) function \cite{Osher1988,Osher2003,Sussman1994}. 

The idea behind the VOF methods is to use the volume fraction (VOF function) of the target fluid in each grid cell and identify the interface by the cells where the volume fraction values fall between 0 and 1. The VOF methods update the VOF function by solving the advection equation with the finite volume scheme, which ensures the mass conservation of the VOF function. 
A class of schemes have been devised so far to explicitly reconstruct the interface with straight or curved lines (plane or curved surface in three dimensions) from the given volume fraction values. For example, in the original VOF scheme \cite{Hirt1981} and the earlier Simple Line Interface Calculation (SLIC) scheme \cite{SLIC1976}, the interface is represented by a straight line parallel to the grid axis, which results in at best the first-order accuracy. As a more accurate algorithm, Piecewise Linear Interface Calculation (PLIC) \cite{Youngs1982, Youngs1984} algorithm was proposed where the orientation (the normal direction) of the cell-wise interface line segment is taken into account. To improve the accuracy of the orientation computation, the least squares volume-of-fluid interface reconstruction algorithm (LVIRA) \cite{LVIRA} and the efficient least squares VOF interface reconstruction algorithm (ELVIRA) \cite{Puckett2004} were developed. Rather than the linear approximation in the SLIC and PLIC schemes, efforts to fit the interface with higher order surface representations have been also made in the past years, for instance, the least-square fit (LSF) \cite{Scardovelli2003,Aulisa2007} algorithm, the spline interface reconstruction (SIR) \cite{Lopez2004} algorithm, piecewise-planar interface reconstruction with cubic-Bezier fit (CBIR) \cite{Lopez2008} algorithm, and the Quadratic Spline based Interface (QUASI) \cite{Diwakar2008} algorithm. Although these algorithms show some improvements compared to the PLIC method, to our knowledge, all these algorithms are limited to 2D and none of them can realize interface reconstruction beyond second-order. The abovementioned VOF methods, known as the geometrical type VOF methods, explicitly use geometrical elements, such as line or surface segments cutting through grid cells, to formulate the interface, thus involve complicated algorithmic procedure which makes the following tasks formidably difficult in 3D case: (1) approximating interface with higher-order surface representation, and (2) implementing for three-dimensional cases. 

In contrast to the geometrical VOF schemes mentioned above, the Tangent of Hyperbola Interface Capturing (THINC) \cite{THINC,THINCSW,MTHINC,UMTHINC,UMTHINC2,THINCQQ} schemes, use an implicit approach to embed the geometrical information in the THINC function, a regularized hyperbolic tangent function, which is Sigmod function mimicking the VOF function. Given the VOF values in all cells, the THINC function can be cell-wisely constructed as continuous function in each cell, which is then used to compute the numerical fluxes needed in the finite volume formulation to update the VOF function. The THINC scheme can be simply used with normal finite volume advection schemes to capture moving interfaces in practice. Some variants of THINC method have developed, including multi-dimensional THINC reconstruction in Cartesian grid \cite{,MTHINC} and unstructured grids in 2 and 3D \cite{UMTHINC,THINCQQ} with linear and quadratic surface representations. It is revealed that the schemes with multi-dimensional THINC reconstructions are able to give numerical results competitive to existing PLIC type VOF methods and show great practical significance and potential for further extension. 

It should be noted that being a discontinuous (or steep-gradient) field function, the VOF function cannot provide accurate and reliable geometrical information for the interface, while the level set method show great superiority as discussed next.

Being another methodology for interface capturing, the level set methods have gained increasing popularity in recent years. Unlike the VOF methods, the level set methods employ the smooth signed distance function to represent the interface. The signed distance function, also known as the LS function, is a set of smooth hypersurfaces or manifolds of higher dimensions, which can be solved by high-order schemes for advection equation. Furthermore, given the signed distance function, the level set methods can provide adequate geometrical information of the interface, such as normal vector and curvature, needed in formulating physical models for interface and reconstructing the interface as explored later in this paper.  
During the evolution of moving interface the values of LS function in the computational domain may not remain always the distance to the interface, so the reinitialization process \cite{Sussman1994,FMM} is needed in practice to modify the  values of the level set function on grid cells. In principle, as the numerical conservativeness is not guaranteed in either advection (evolution) or reinitialization computation, mass loss or gain may occur to the fluid of interest, which is the fatal drawback of level set method. Several efforts have been made to remedy the conservation error for level set methods, for example, the coupled particle/level set  (Particle LS) \cite{Enright2002} and the coupled the level set/VOF methods (CLSVOF) \cite{Sussman2000,Menard2007, ACLSVOF, VOSET}. Unfortunately, these methods are of significant algorithmic complexity, and none of them realizes both numerical conservativeness and high-order interface representation.

In this paper, we present a novel interface capturing scheme, so-called THINC/LS (THINC scheme coupled with the LS method), which possesses the following major new properties in comparison with other state-of-the-art methods. 
\begin{itemize}
\item Instead of the plane and quadratic surface representations used in the existing geometrical VOF methods, higher-order (arbitrary order in principle) surface polynomials can be determined from the level set function, and then used to build the THINC or VOF function. 
\item The interface in the level set field, i.e. the 0-level surface in the level set function, is synchronized with the interface identified from the THINC reconstruction with the mass conservation constraint. Thus, the numerical conservativeness is ensured. 
\item The function pair of THINC and level set are updated and coupled simultaneously through simple and straightforward solution procedure without the complex computations involved in the geometrical VOF algorithms. 
\end{itemize}

We have extensively examined the THINC/LS method regarding interface reconstruction and transport of moving interfaces. The numerical results verify the superiority of the method in comparison with other existing methods. 

This paper is organized as follows. The details of the proposed THINC/LS scheme are described in section 2. Interface reconstruction tests are presented in section 3. Extensive numerical verifications with widely used benchmark tests are presented in section 4. We end this paper with some conclusion remarks in section 5.

\section{Numerical method}

\subsection{The interface indication functions}

Although the idea presented in this work can be in principle extended to any structured and unstructured mesh, we limit our presentation to Cartesian grid. The computational domain is divided by mesh cell $\Omega_{ijk}=[x_{i-\frac{1}{2}},x_{i+\frac{1}{2}}] \times [y_{j-\frac{1}{2}},y_{j+\frac{1}{2}}] \times [z_{k-\frac{1}{2}},z_{k+\frac{1}{2}}]$, where $i$, $j$ and $k$ denote the cell index in $x$, $y$ and $z$ directions respectively. 
We also denote the six surface segments of $\Omega_{ijk}$ by 
\begin{equation}
\begin{array}{l}
 S_{i\pm\frac{1}{2}jk}: x_{i\pm\frac{1}{2}}\times [y_{j-\frac{1}{2}}, y_{j+\frac{1}{2}}] \times[z_{k-\frac{1}{2}}, z_{k+\frac{1}{2}}], \\
 S_{ij\pm\frac{1}{2}k}: [x_{i-\frac{1}{2}},x_{i+\frac{1}{2}}] \times y_{j\pm\frac{1}{2}} \times[z_{k-\frac{1}{2}}, z_{k+\frac{1}{2}}], \\
 S_{ijk\pm\frac{1}{2}}: [x_{i-\frac{1}{2}},x_{i+\frac{1}{2}}] \times [y_{j-\frac{1}{2}}, y_{j+\frac{1}{2}}] \times z_{k\pm\frac{1}{2}}. 
\end{array}
\end{equation}

In the present work,  we use a mesh of uniform spacing ($\Delta x=x_{i+\frac{1}{2}}-x_{i-\frac{1}{2}}=\Delta y=y_{j+\frac{1}{2}},-y_{j-\frac{1}{2}}=\Delta z=z_{k+\frac{1}{2}}-z_{k-\frac{1}{2}}$) for simplicity. The center of each cell locates at $(x_i,y_j,z_k)$. 

Considering an interface $\partial V$ lying between two regions, $V^1$ and $V^2$ filled with different fluids in $\mathbf{R}^2$ or $\mathbf{R}^3$,
 we introduce two interface indicator functions, i.e. the level set (LS)  function $\phi(\textbf{x},t)$ and the THINC function $H(\textbf{x},t)$, as illustrated in  \autoref{fun} for the one-dimensional case. 
\begin{itemize}
\item LS function (\autoref{fun}(a)): 

The distance from a point $\textbf{x}=(x,y,z)$ to the interface $\partial V$ is defined by
\begin{equation}
\text{dist}(\textbf{x},\partial V)=\inf_{\textbf{x}_I\in \partial V}\|\textbf{x}-\textbf{x}_I\|,
\end{equation}

where $\textbf{x}_I$ represents any  point on the interface, and is called the interface point hereafter.

The LS function is a signed distance function to the interface, given by
\begin{equation}
\phi(\textbf{x},t)=\left\{
\begin{array}{rcc}
\text{dist}(\textbf{x},\partial V) & & \text{if } \textbf{x}\in V^1 \\
0\quad\quad\quad & & \text{if } \textbf{x}\in\partial V \\
-\text{dist}(\textbf{x},\partial V) & & \text{if } \textbf{x}\in V^2
\end{array}, \right.
\end{equation}

where $\partial V$ denotes the interface, $V^1$ the region for fluid 1 and $V^2$ the region for fluid 2.  The LS function is differentiable and facilitates accurate computations for the geometrical properties of the interface.

\begin{figure}[htbp]
	\centering
	\subfigure[] {
		\centering
		\includegraphics[width=0.35\textwidth]{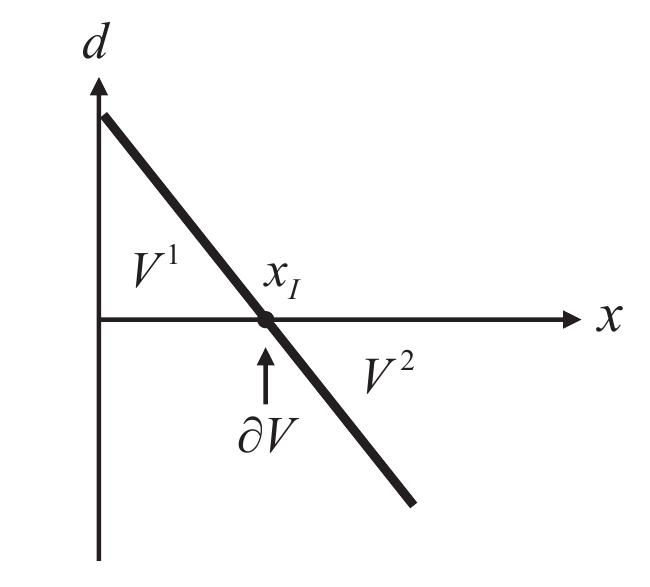}} \hspace{2.0cm}
	\subfigure[] {
		\centering
		\includegraphics[width=0.35\textwidth]{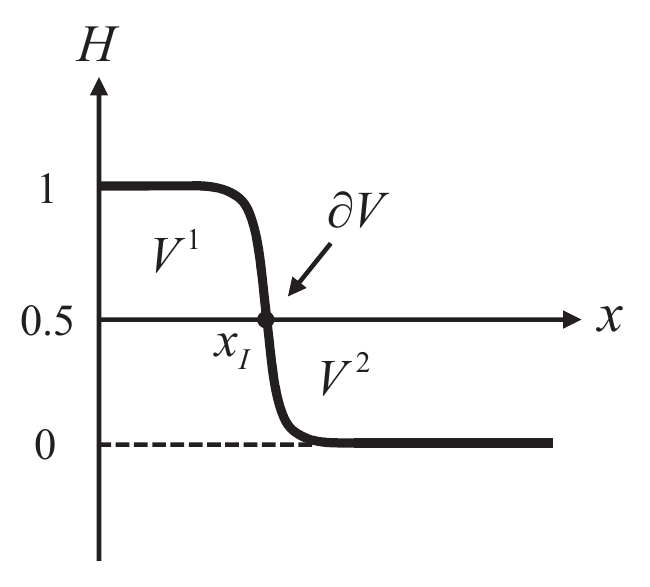}}
	\caption{Schematic diagram of the interface indicator functions in one-dimension: (a) The LS function; (b) The THINC function.}
\label{fun}
\end{figure}

\item THINC function (\autoref{fun}(b)): 

The THINC function is designed to represent a jump-like solution using the hyperbolic tangent function, 
\begin{equation}
H(\textbf{x},t)=\frac{1}{2}\left(1+\tanh\left(\beta \phi(\textbf{x},t)\right) \right),  
\label{h-def1}
\end{equation}
where $\beta$ is a parameter to control the thickness of the interface, i.e. the width of the jump in the hyperbolic tangent function. See a more detailed discussion on $\beta$ in \cite{THINCSW}. 

Being essentially a continuous function, the THINC function is able to approximate the Heavisde step function by increasing the value of $\beta$.  The cell-average of THINC function 
\begin{equation}
\bar{H}_{ijk}(t)=\frac{1}{|\Omega_{ijk}|}\int_{\Omega_{ijk}}H(\textbf{x},t)dxdydz
\label{vof-def}
\end{equation}
stands for the volume fraction of a specified fluid in the VOF method. Thus, the rigorous conservation of mass/volume can be automatically guaranteed  if the finite volume formulation is applied to solve the THINC function as in the normal VOF schemes. 
\end{itemize}

From the above definitions, the interface can be equivalently identified by either $\phi(\textbf{x},t)=0$ or $H(\textbf{x},t)=\frac{1}{2}$. 

The proposed THINC/LS method updates the LS function $\phi(\textbf{x},t)$ and the THINC function $H(\textbf{x},t)$ simultaneously and synchronizes them to realize both high-order interface representation and numerical conservativeness.

\subsection{The evolution equations}

In the present work, we assume that the free interface is passively transported by fluid motion. The advection equation of the LS function $\phi(\textbf{x},t)$ reads
\begin{equation}
\frac{\partial \phi}{\partial t}+\textbf{u}\cdot \nabla \phi=0,
\label{ls-adv}
\end{equation}
where $\textbf{u}=(u,v,w)$ is the flow velocity.

The THINC function $H(\textbf{x},t)$ follows the same advection equation,  
\begin{equation}
\frac{\partial H}{\partial t}+\textbf{u}\cdot \nabla H=0.
\label{h-adv}
\end{equation}

In order to implement the finite volume method, we rewrite \eqref{h-adv} into the flux form as
\begin{equation}
\frac{\partial H}{\partial t}+\nabla\cdot(\textbf{u}H)-H\nabla\cdot\textbf{u}=0.
\label{h-flux}
\end{equation}
It is obvious that \eqref{h-flux} reduces to 
\begin{equation}
\frac{\partial H}{\partial t}+\nabla\cdot(\textbf{u}H)=0
\end{equation}
for incompressible flows. Shown latter, the cell-average of THINC function 
$\bar{H}_{ijk}(t)$, i.e the volume fraction or VOF function defined in \eqref{vof-def},  is updated via the finite volume method so as to ensure rigorous numerical conservation.

\subsection{The THINC/LS method}
 As the central part of this paper, we describe the numerical algorithm of the THINC/LS method in this subsection.

We suppose that at time step $n$ ($t=t^n$), we are given with the VOF values $\bar{H}^n_{ijk}$ of the interface cells $\Omega_{ijk}$ where the interface of interest falls in, as well as the values of LS function, $\phi^n_{i'j'k'}$,  at the centers of the supporting cells  $\Omega_{i'j'k'}$, encompassing  $\Omega_{ijk}$, $\Omega_{ijk}\in\cup\Omega_{i'j'k'}$.  

\subsubsection{Retrieve the high-order surface polynomial from LS function}

We approximate the cell-wise LS function within cell $\Omega^n_{ijk}$ by 
\begin{equation}
\begin{split}
\mathcal{P}_{ijk}(x,y,z) = \sum_{\substack{
{s,t,r=0}}
}^{p}a_{str}X^sY^tZ^r,
\end{split}
\label{s-poly}
\end{equation}
where $\mathcal{P}_{ijk}$ is a polynomial of $p$th order and $(X, Y, Z)$ is the local coordinates for cell $\Omega_{ijk}$, i.e.  
$(X,Y,Z)=(x-x_i,y-y_j,z-z_k)$. $\mathcal{P}_{ijk}(x,y,z)=0$,  referred to as the surface polynomial in our context, represents the 0-level surface of level set function cutting through the target cell. The coefficients $a_{str}$ are computed from the supporting cells  $\Omega_{i'j'k'}$ using Lagrange interpolation, 
\begin{equation}
\mathcal{P}_{ijk}(x_{i'_l},y_{j'_l},z_{k'_l}) = \phi^n_{i'_lj'_lk'_l}, \quad l \in {0,1,...,p}.
\label{s-cell}
\end{equation}

Substitute \eqref{s-cell} into \eqref{s-poly}, we get a system of linear equations to find  the coefficients $a_{str}$. The system in matrix-vector form reads
\begin{equation}
\begin{bmatrix}
1 & X_{i'_0} & Y_{j'_0} & Z_{k'_0}& ...& X^p_{i'_0}Y^p_{j'_0}Z^p_{k'_0}\\
1 & X_{i'_1} & Y_{j'_0} & Z_{k'_0}& ...& X^p_{i'_1}Y^p_{j'_0}Z^p_{k'_0}\\
\vdots & \vdots & \vdots & \vdots & & \vdots\\
1 & X_{i'_p} & Y_{j'_p} & Z_{k'_p}& ...& X^p_{i'_p}Y^p_{j'_p}Z^p_{k'_p}
\end{bmatrix}
\begin{bmatrix}
a_{000}\\
a_{100}\\
\vdots\\
a_{ppp}
\end{bmatrix}
=\begin{bmatrix}
\phi^n_{i'_0j'_0k'_0}\\
\phi^n_{i'_1j'_0k'_0}\\
\vdots\\
\phi^n_{i'_pj'_pk'_p}
\end{bmatrix}.
\label{matrix-ploy}
\end{equation}

Considering the symmetry of the supporting cells, polynomials of even order $(p=2,4)$ are considered in this paper.

\subsubsection{Enforce mass/volume conservation from the volume fraction constraint}
In the solution procedure of the present  THINC/LS  algorithm shown latter, the LS values $\phi^n_{ijk}$ are obtained by transporting the LS function with the advection equation, thus the interface defined by $\phi(x,y,z,t^n)=0$ does not guarantee mass/volume conservation. 
We use the volume fraction $\bar{H}^n_{ijk}$ as the constraint condition to enforce the conservation. 

Given the surface polynomial $\mathcal{P}_{ijk}$ computed by \eqref{s-poly}, as well as the volume fraction $\bar{H}^n_{ijk}$, we write the THINC function \eqref{h-def1} as,    
\begin{equation}
H(x,y,z,t)=\frac{1}{2}\left(1+\tanh\left(\beta \left( \mathcal{P}_{ijk}(x,y,z)+\phi^{\bigtriangleup}_{ijk}\right)\right) \right),  
\label{h-def2}
\end{equation}
where  $\phi^{\bigtriangleup}_{ijk}$ is the correction to the level set function due to the volume fraction constraint, which is determined so that 
\begin{equation}
\frac{1}{|\Omega_{ijk}|}\int_{\Omega_{ijk}}\frac{1}{2}\left(1+\tanh\left(\beta \left( \mathcal{P}_{ijk}(x,y,z)+\phi^{\bigtriangleup}_{ijk}\right)\right) \right)=\bar{H}^n_{ijk}
\label{mass-constraint}
\end{equation}
is satisfied. 

Following \cite{THINCQQ},  we use Gaussian quadrature to approximate the spatial integration in \eqref{mass-constraint},  which yields
\begin{equation}
\sum_{g=1}^Gw_g\frac{1}{2}\left(1+\tanh\left(\beta \left( \mathcal{P}_{ijk}(x,y,z)+\phi^{\bigtriangleup}_{ijk}\right)\right) \right)=\bar{H}^n_{ijk}, 
\label{g-quadrature}
\end{equation}
where $w_g$ is the normalized  weights at  Gaussian quadrature points satisfying $\sum_{g=1}^Gw_g=1$. The number of quadrature points should be chosen according to the desired numerical accuracy.  See \cite{THINCQQ} for more discussions. In the present work, we use 3 Gaussian points in each dimension. Newton iteration method can be used to solve \eqref{g-quadrature}. Using 0 as the first guess, the iteration converges within a few steps. As long as $\phi^{\bigtriangleup}_{ijk}$ is obtained, the interface segment in cell 
$\Omega_{ijk}$ can be identified as 
\begin{equation}
 \mathcal{P}_{ijk}(x,y,z)+\phi^{\bigtriangleup}_{ijk}=0. 
\label{eq-interface}
\end{equation}

\begin{figure}[htbp]
	\centering
	\includegraphics[width=0.4\textwidth]{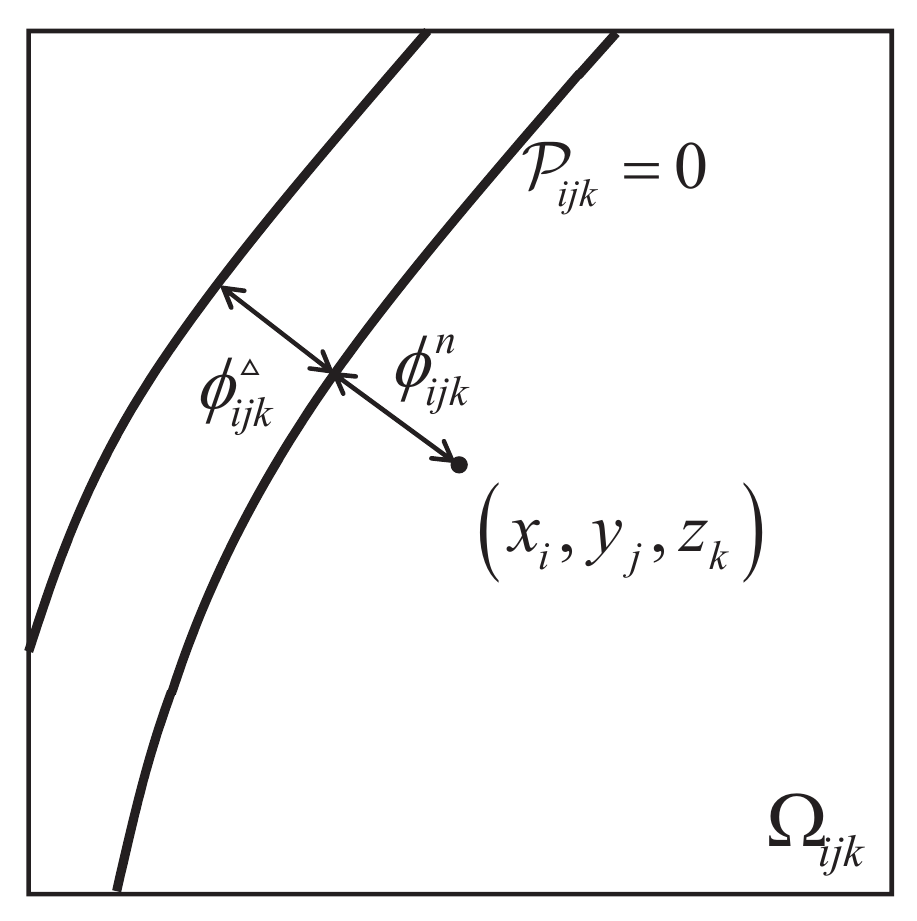}
	\caption{Schematic diagram of the surface polynomial and the volume fraction constraint.}
\label{LS-poly}
\end{figure}

Remarks: we make use of the LS function to find the cell-wise surface polynomial $\mathcal{P}_{ijk}$ that includes all geometrical information of the interface. The surface polynomial $\mathcal{P}_{ijk}$ can be viewed as a local LS function for the target cell. However, the interface represented by 
\begin{equation}
\mathcal{P}_{ijk}=0
\label{ls-1} 
\end{equation}
does not necessarily satisfy the mass/volume conservation. As shown in \autoref{LS-poly}, using the constraint condition \eqref{mass-constraint}, we adjust (or synchronize) a distance of $\phi^{\bigtriangleup}_{ijk}$ in the normal direction of the interface so as to enforce the conservation.
  
\subsubsection{Reinitialization}

After $\phi^{\bigtriangleup}$ are obtained for all interface cells, we need to reinitialize $\phi^{n}$. The reinitialization process is equivalent to solving the Eikonal equation,
\begin{equation}
|\nabla \phi|=1
\label{reinit}
\end{equation}
with the VOF values $\phi^{nc}_{ijk}=\phi^{n}_{ijk}+\phi^{\bigtriangleup}_{ijk}$ fixed in the interface cells, where $|\phi^{nc}_{ijk}|=\le \Delta x$. 

In the present work, we use the Fast Sweeping Method (FSM) \cite{FSM} for reinitialization. The LS function after reinitialization \eqref{reinit},  $\phi^{nc}$, satisfies the mass/volume conservation condition. 

\subsubsection{Update the level set and THINC functions}

Given the mass conserving LS function $\phi^{nc}$ at time step $n$ ($t^{n}$), we solve the advection evolution \eqref{ls-adv} to update the LS function to step $n+1$ ($t^{n+1}=t^{n}+\Delta t$),  $\phi^{n+1}$.   In the present work, we use the fifth-order Hamilton-Jacobi WENO scheme \cite{HJWENO} for spatial discretization and the third-order TVD Runge-Kutta scheme \cite{TVDRK3,SSPRK} for time integration. It it noted that  $\phi^{n+1}$ computed from the advection equation does not necessarily satisfy the mass/volume conservation. 

The cell-average values of the THINC function, i.e. the volume fractions  are updated via the finite volume formulation as follows 
\begin{equation}
\begin{split}
\frac{d {\bar{H}_{ijk}(t)}}{d t} & =-\left(\frac{1}{\Delta x}\left(\mathcal{F}^{<x>}_{i+\frac{1}{2}jk}-\mathcal{F}^{<x>}_{i-\frac{1}{2}jk}\right )+\frac{1}{\Delta y}\left(\mathcal{F}^{<y>}_{ij+\frac{1}{2}k}-\mathcal{F}^{<y>}_{ij-\frac{1}{2}k}\right )+\frac{1}{\Delta z}\left(\mathcal{F}^{<z>}_{ijk+\frac{1}{2}}-\mathcal{F}^{<z>}_{ijk-\frac{1}{2}}\right )\right ),
\end{split}
\label{dhdt}
\end{equation}
where $\mathcal{F}^{<\alpha>}, \ \alpha=x,y,z$, denote the numerical fluxes  in $x,y,z$ directions respectively, which are computed over each cell surface by Gaussian quadrature. We show the numerical formula for $\mathcal{F}^{<x>}_{i+\frac{1}{2}jk}$, the numerical flux in $x$ direction through cell surface segment $S_{{i+\frac{1}{2}jk}}$, as an example,  
\begin{equation}
\mathcal{F}^{<x>}_{i+\frac{1}{2}jk}=\sum_{g=1}^Gw_g\left (u(x_{i+\frac{1}{2}},y_g,z_g) H_{i^{up}jk}(x_{i+\frac{1}{2}},y_g,z_g) \right), 
\label{x-flux}
\end{equation}
where $(y_g,z_g)$ stands for the Gaussian points on cell surface 
$S_{{i+\frac{1}{2}jk}}$. The subscript $i^{up}$ indicates the THINC reconstruction function in the upwinding cell which is determined by
\begin{equation}
i^{up}=\left\{
\begin{array}{lll}
i, & & {\rm if} \ u(x_{i+\frac{1}{2}},y_g,z_g)>0; \\
i+1, & & \rm{otherwise.}
\end{array} \right.
\end{equation}
Given the THINC function \eqref{h-def2} and the velocity specified at the Gauss points, we get the numerical flux \eqref{x-flux} immediately. The same applies to numerical fluxes on other surface segments of $\Omega_{ijk}$. 

After the numerical fluxes are obtained, the third-order TVD Runge-Kutta scheme is used to updated to the volume fraction to $n+1$ time step ($\bar{H}^{n+1}_{ijk}$). 

Note that the THINC reconstruction and the computation to update the volume fraction  are limited only to the interface cells identified by $\epsilon\le \bar{H}^{n}_{ijk} \le 1-\epsilon$. $\epsilon$ is a small positive and set to $10^{-8}$ in the present work.

\subsubsection{Summary of the solution procedure}
In order to facilitate the implementation of the present method, we summarize the computation steps to update the function pair from ($\phi^{n} , \bar{H}^{n})$ to   ($\phi^{n+1} , \bar{H}^{n+1})$ as follows. 

\begin{description}
\item {Step 1: } Calculate the surface polynomial $\mathcal{P}_{ijk}$ in \eqref{s-poly} for interface cells $\epsilon\le \bar{H}^{n}_{ijk} \le 1-\epsilon$ with the coefficients computed from \eqref{matrix-ploy} using the values of LS function $\phi^n$ in the supporting cells; 

\item {Step 2: } Compute volume fraction constraint $\phi^{\bigtriangleup}_{ijk} $ for interface cells $\epsilon\le \bar{H}^{n}_{ijk} \le 1-\epsilon$ that satisfies \eqref{mass-constraint} from the Gaussian quadrature  formula \eqref{g-quadrature} with the given volume fractions $\bar{H}^{n}$;

\item {Step 3: } Take  the interface cells $|\phi^{nc}_{ijk}=\phi^{n}_{ijk} +\phi^{\bigtriangleup}_{ijk} |\le \Delta x$ as the new 0-level set, and reinitialize the LS values for other cells, which synchronize the level function  $\phi^{n}_{ijk} $ to  $\phi^{nc}_{ijk} $ to fulfill the mass/volume conservation; 

\item {Step 4: } Update the level set function from $\phi^{nc}$ to $\phi^{n+1}$ by the Runge-Kutta time integration scheme and finite difference spatial discretization; 

\item {Step 5: } Update the volume fraction by \eqref{dhdt} where the numerical fluxes of interface cells $\epsilon\le \bar{H}^{n}_{ijk} \le 1-\epsilon$ are computed from the THINC function \eqref{h-def2} using Gaussian quadrature. The Runge-Kutta is used to predict the volume fraction $\bar{H}^{n+1}$ to new time level.  

\item {Step 6: } Go back to step 1 to repeat the computations for next time level.   

\end{description}

\section{Interface reconstruction tests}

We now assess the interface reconstruction accuracy and convergence rates of the proposed THINC/LS method with some known geometries. In order to quantify the numerical accuracy and convergence rates, we identify the reconstruction error $E_r$ as the difference between the area (or volume in 3D)  encompassed by the numerical  reconstructed interface and the exact interface \cite{Rider1998}
\begin{equation}
E_r=\sum_{ijk}\iiint_{\Omega_{ijk}}|H(x,y,z)-H^0(x,y,z)|dxdydz,
\label{err_rec}
\end{equation}
where $H(x,y,z)$ and $H^0(x,y,z)$ denote the numerical and exact solutions  respectively to the VOF function.

\subsection{Circle reconstruction}
\begin{figure}[htbp]
	\centering
	\subfigure[] {
		\centering
		\includegraphics[width=0.45\textwidth]{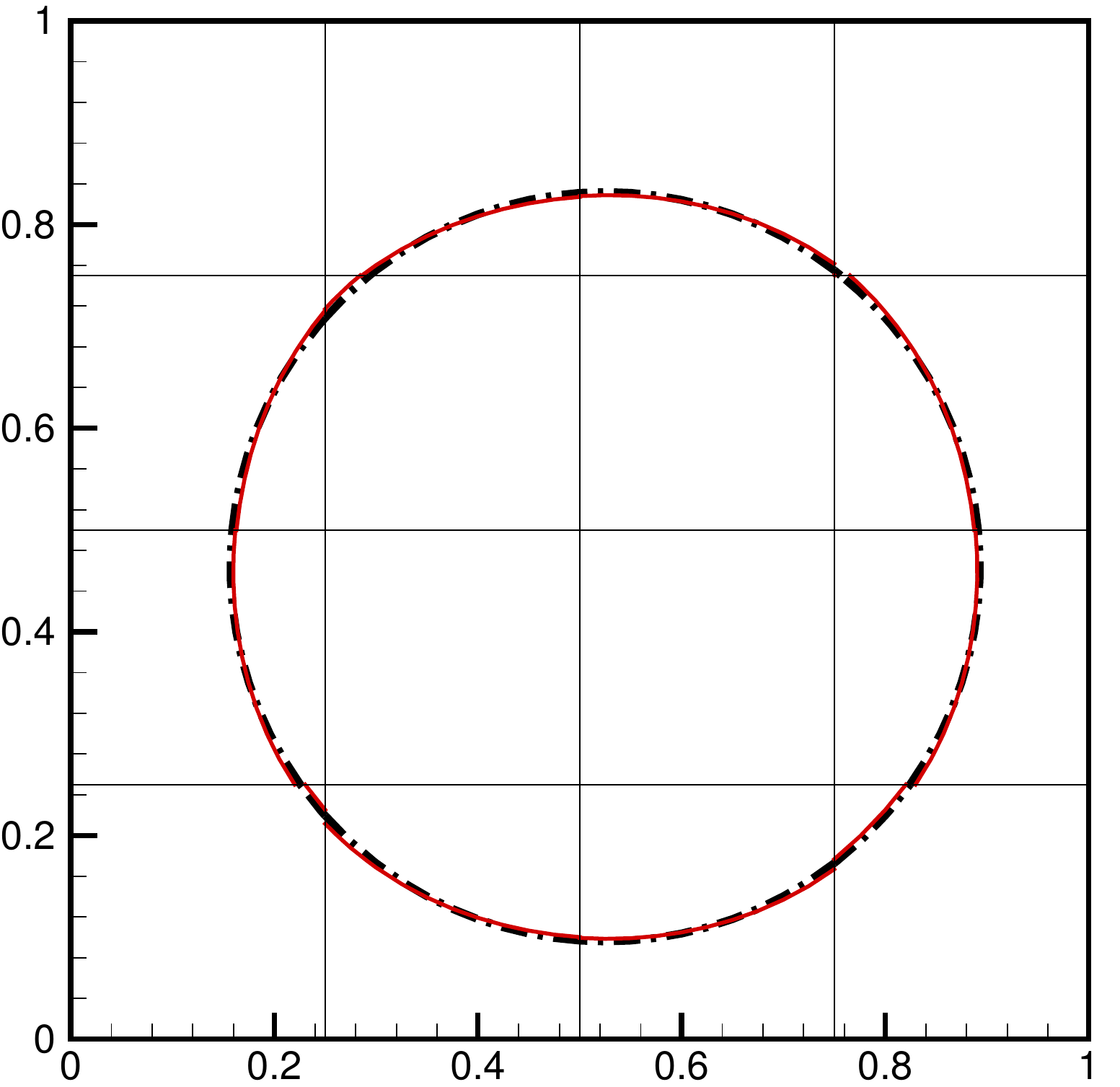} }\hspace{1cm}
	\subfigure[] {
		\centering
		\includegraphics[width=0.45\textwidth]{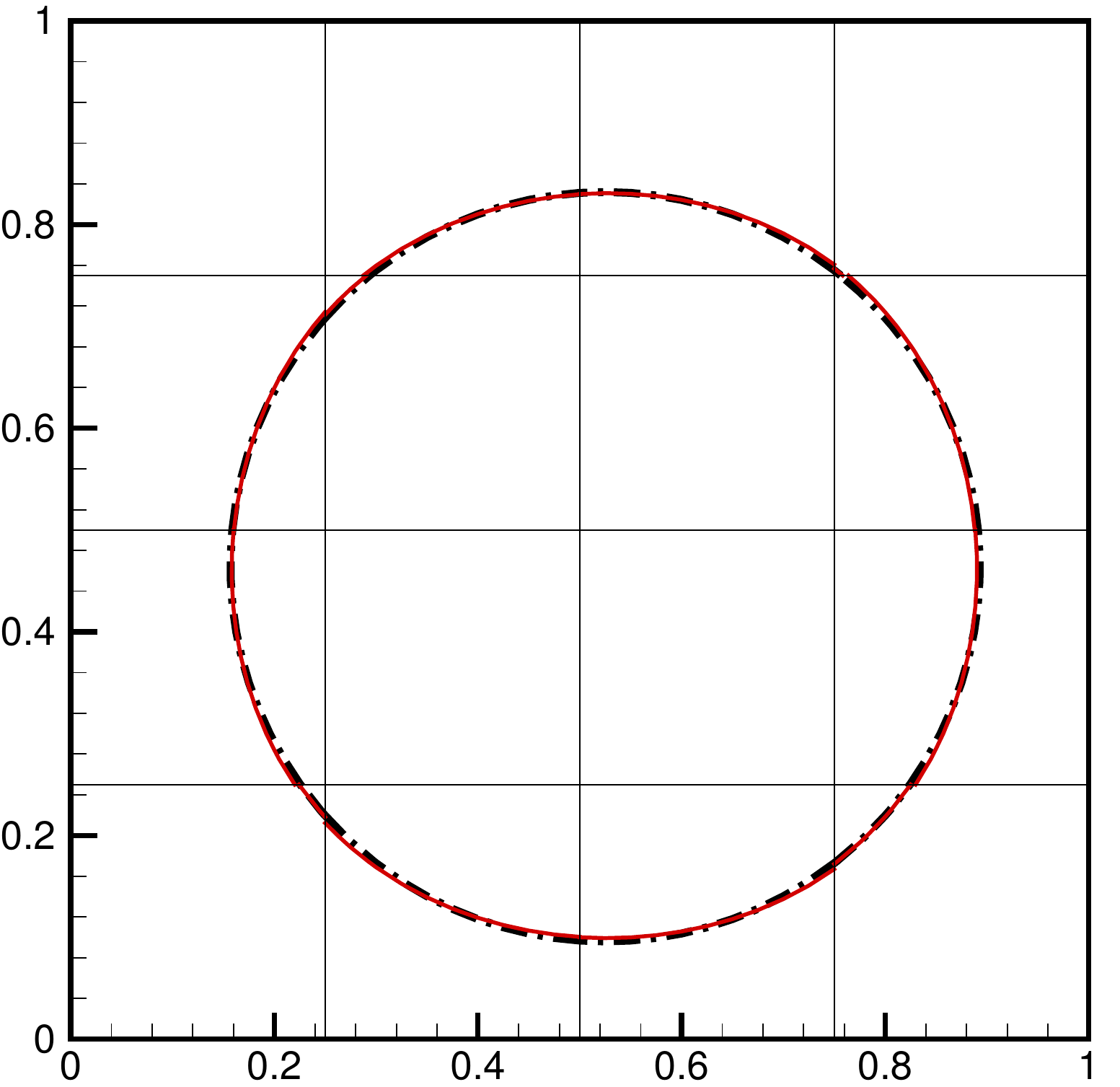} }\hspace{1cm}
	\caption{Reconstructed circle curves using (a) second-order ($p=2$) and (b) fourth-order ($p=4$) polynomials on a $4 \times 4$ mesh. Black dash dot line is the exact solution and red solid line is the numerical solution. Both lines are generated of the VOF values sampled at the $10 \times 10$ uniformly distributed sample points within each mesh cell.}
\label{recon-circle-f}
\end{figure}

As described in \cite{Rider1998}, a circle with the radius of 0.368 is centered at $(0.525, 0.464)$ in a unit domain. We use the second-order ($p=2$) and fourth-order ($p=4$) polynomials to represent the interface. The reconstructed curves are illustrated in \autoref{recon-circle-f}. It is fount that both curves coincide well with the exact solution even on a $4\times 4$ mesh. The fourth-order polynomial gives more accurate results than the second-order one.

For a detailed comparison of different order polynomials, we list the numerical errors and convergence rates in \autoref{recon-circle-t} and compare them with some classical geometrical VOF methods. As shown in \autoref{recon-circle-t}, the numerical errors and convergence rates of the THINC/LS method are much better than the Piecewise Linear Interface Calculation (PLIC) VOF methods. We observe the high-order convergence rates as expected, which implies that the THINC/LS method can achieve arbitrary order interface reconstruction given the high-order surface polynomial.

\begin{sidewaystable}[htbp]

	\begin{center}
    \protect\caption{$L_1$ Circle Reconstruction Error Norms and convergence rates for different methods  $(\beta=3.5/\Delta x)$}
    \label{recon-circle-t}
		\begin{tabular}{lccccccccccc}
		\hline
Methods	& $10^2$ &	Order	& $20^2$ &	Order	&$40^2$ &	Order	& $80^2$ &	Order	& $160^2$ &	Order	& $320^2$\\
		\hline
Youngs \cite{Lopez2004} & $2.09\times10^{-3}$ & 1.79 & $6.06\times10^{-4}$ & 1.18 & $2.67\times10^{-4}$ & 1.31 & $1.08\times10^{-4}$ & 1.03 & $5.29\times10^{-5}$ & - & -\\
CD \cite{Lopez2004} & $2.34\times10^{-3}$ & 2.06 & $5.61\times10^{-4}$ & 1.89 & $1.51\times10^{-4}$ & 1.88 & $4.10\times10^{-5}$ & 1.81 & $1.17\times10^{-5}$ & - & -\\
Puckett \cite{Lopez2004} & $3.36\times10^{-3}$ & 2.27 & $6.96\times10^{-4}$ & 2.05 & $1.68\times10^{-4}$ & 1.94 & $4.38\times10^{-5}$ & 1.95 & $1.13\times10^{-5}$ & - & -\\
ELVIRA \cite{Lopez2004} & $2.94\times10^{-3}$ & 2.24 & $6.22\times10^{-4}$ & 1.98 & $1.58\times10^{-4}$ & 1.93 & $4.15\times10^{-5}$ & 2.05 & $1.00\times10^{-5}$ & - & -\\
SIR \cite{Lopez2004} & $1.94\times10^{-3}$ & 2.03 & $4.76\times10^{-4}$ & 1.88 & $1.29\times10^{-4}$ & 2.15 & $2.91\times10^{-5}$ & 1.89 & $7.86\times10^{-6}$ & - & -\\
	       \hline 
THINC/LS (p=2) & $4.27\times10^{-4}$ & 3.02 & $5.26\times10^{-5}$ & 2.99 & $6.64\times10^{-6}$ & 3.01 & $8.25\times10^{-7}$ & 3.00 & $1.03\times10^{-7}$ & 3.01  & $1.28\times10^{-8}$ \\	
THINC/LS (p=4) & $6.28\times10^{-5}$ & 5.08 & $1.85\times10^{-6}$ & 4.99 & $5.84\times10^{-8}$ & 5.00 & $1.83\times10^{-9}$ & 5.00 & $5.72\times10^{-11}$  & 5.00  & $1.78\times10^{-12}$\\	       
	       \hline       	        
		\end{tabular}
	\end{center}

\bigskip\bigskip\bigskip

	\begin{center}
    \protect\caption{$L_1$ Sphere Reconstruction Error Norms and convergence rates for different methods $(\beta=3.5/\Delta x)$} 
\label{recon-sphere-t}
		\begin{tabular}{lccccccccccc}
		\hline
Methods	& $10^2$ &	Order	& $20^2$ &	Order	&$40^2$ &	Order	& $80^2$ &	Order	& $160^2$ &	Order	& $320^2$\\
		\hline
Youngs \cite{Lopez2008} & $1.89\times10^{-3}$ & 1.84 & $5.28\times10^{-4}$ & 1.45 & $1.93\times10^{-4}$ & 1.17 & $8.60\times10^{-5}$ & 1.06 & $4.12\times10^{-5}$ & 1.02 & $2.03\times10^{-5}$\\
LSF \cite{Aulisa2007}  & $1.92\times10^{-3}$ & 2.01 & $4.77\times10^{-4}$ & 2.00 & $1.19\times10^{-4}$ & 2.00 & $2.98\times10^{-5}$ & 2.00 & $7.46\times10^{-6}$ & - & -\\
LLCIR \cite{Lopez2008}  & $2.14\times10^{-3}$ & 2.03 & $5.23\times10^{-4}$ & 1.89 & $1.41\times10^{-4}$ & 1.61 & $4.61\times10^{-5}$ & 1.32 & $1.85\times10^{-5}$ & 1.14 & $8.40\times10^{-6}$\\
ELCIR \cite{Lopez2008}  & $2.23\times10^{-3}$ & 1.99 & $5.62\times10^{-4}$ & 1.98 & $1.42\times10^{-4}$ & 1.88 & $3.86\times10^{-5}$ & 1.69 & $1.20\times10^{-5}$ & 1.42 & $4.47\times10^{-6}$\\
CLCIR \cite{Lopez2004}  & $2.38\times10^{-3}$ & 2.11 & $5.50\times10^{-4}$ & 2.08 & $1.30\times10^{-4}$ & 2.01 & $3.23\times10^{-5}$ & 2.01 & $8.00\times10^{-6}$ & 2.00 & $2.00\times10^{-6}$\\
CBIR \cite{Lopez2008}  & $2.43\times10^{-3}$ & 2.11 & $5.64\times10^{-4}$ & 2.12 & $1.30\times10^{-4}$ & 2.03 & $3.18\times10^{-5}$ & 2.02 & $7.82\times10^{-6}$ & 2.00& $1.95\times10^{-6}$\\
	       \hline 
THINC/LS (p=2) & $2.02\times10^{-4}$ & 3.09 & $2.38\times10^{-5}$ & 2.99 & $3.00\times10^{-6}$ & 3.00 & $3.75\times10^{-7}$ & 3.00 & $4.68\times10^{-8}$  & 3.00  & $5.84\times10^{-9}$\\	      
THINC/LS (p=4) & $4.78\times10^{-5}$ & 7.56 & $2.53\times10^{-7}$ & 6.35 & $3.10\times10^{-9}$ & 5.23 & $8.25\times10^{-11}$ & 5.01 & $2.56\times10^{-12}$  & 5.00  & $8.01\times10^{-14}$\\	      		       
	       \hline       	        
		\end{tabular}
	\end{center}
\end{sidewaystable}

\subsection{Sphere reconstruction}
\begin{figure}[htbp]
	\centering
	\subfigure[] {
		\centering
		\includegraphics[width=0.4\textwidth]{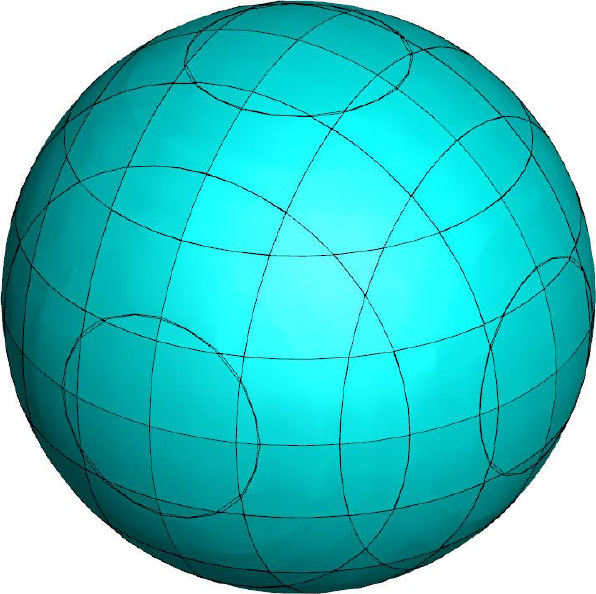} }\hspace{1cm}
	\subfigure[] {
		\centering
		\includegraphics[width=0.4\textwidth]{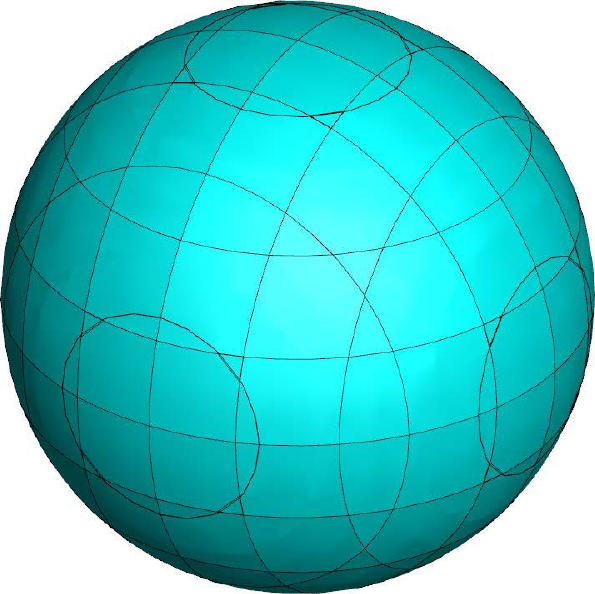} }\hspace{1cm}
	\caption{Reconstructed sphere surfaces using (a) second-order ($p=2$) and (b) fourth-order ($p=4$) polynomials on a $10^3$ mesh. The surfaces are interpolated from the VOF values at $5^3$ uniformly distributed sample points within each cell. The curve lines indicate the edges of the exact and reconstructed surface segments. }
\label{recon-sphere-f}
\end{figure}

For the three-dimensional case, we conduct the reconstruction test presented in \citep{Aulisa2007}. A sphere is centred at $(0.5, 0.5, 0.5)$ in a unit cube with radius 0.325. We use different order ($p=2,4$)  polynomials to reconstruct the surface segments for the interface cells as shown in \autoref{recon-sphere-f}. It is observed that both 2nd and 4th order polynomial surfaces produce accurate results the sphere, and 4th-order polynomial surface coincides better with the exact one.

To compare the THINC/LS method quantitatively against other geometrical VOF methods, we list the reconstruction errors and convergence rates on different mesh resolutions in \autoref{recon-sphere-t}. It is observed that the THINC/LS method produces more accurate results in than the geometrical VOF methods where low-order surface polynomials were used. It is noted that unlike the geometrical VOF methods which are of significant algorithmic complexity, the interface reconstruction procedure in THINC/LS method simple and straightforwardly extendable to even higher-order in both 2 and 3D.

\section{Advection tests}

In this section, we evaluated the proposed THINC/LS method as an interface-capturing scheme with some widely used advection benchmark tests. To quantify the accuracy of the numerical results, we define the $L_1$  error norm for VOF field as
\begin{equation}
E(L_1)=\sum_{ijk}|\bar{H}_{ijk}-\bar{H}_{ijk}^{0}||\Omega_{ijk}|,
\end{equation}
where $\bar{H}_{ijk}$ and $\bar{H}^0_{ijk}$ stand for the numerical and exact VOF function respectively.

Another widely used measure of the VOF methods is the relative error defined by
\begin{equation}
E_{r}=\frac{\sum_{ijk}|H_{ijk}-H_{ijk}^0|}{\sum_{ijk}H_{ijk}^0}.
\end{equation}

\subsection{Translation test}

\begin{figure}[htbp]
	\centering
	\subfigure[] {
		\centering
		\includegraphics[width=0.3\textwidth]{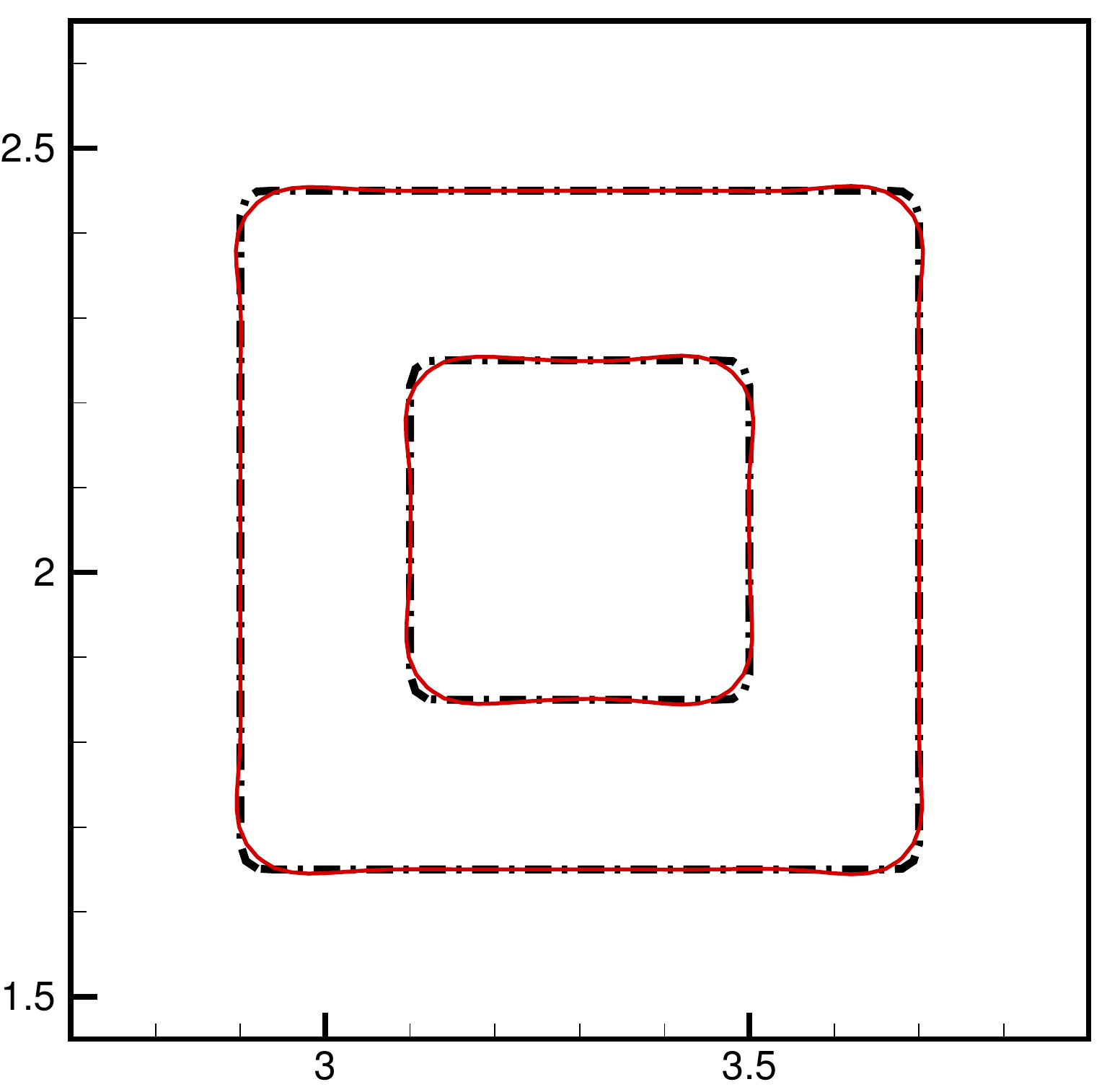} }
	\subfigure[] {
		\centering
		\includegraphics[width=0.3\textwidth]{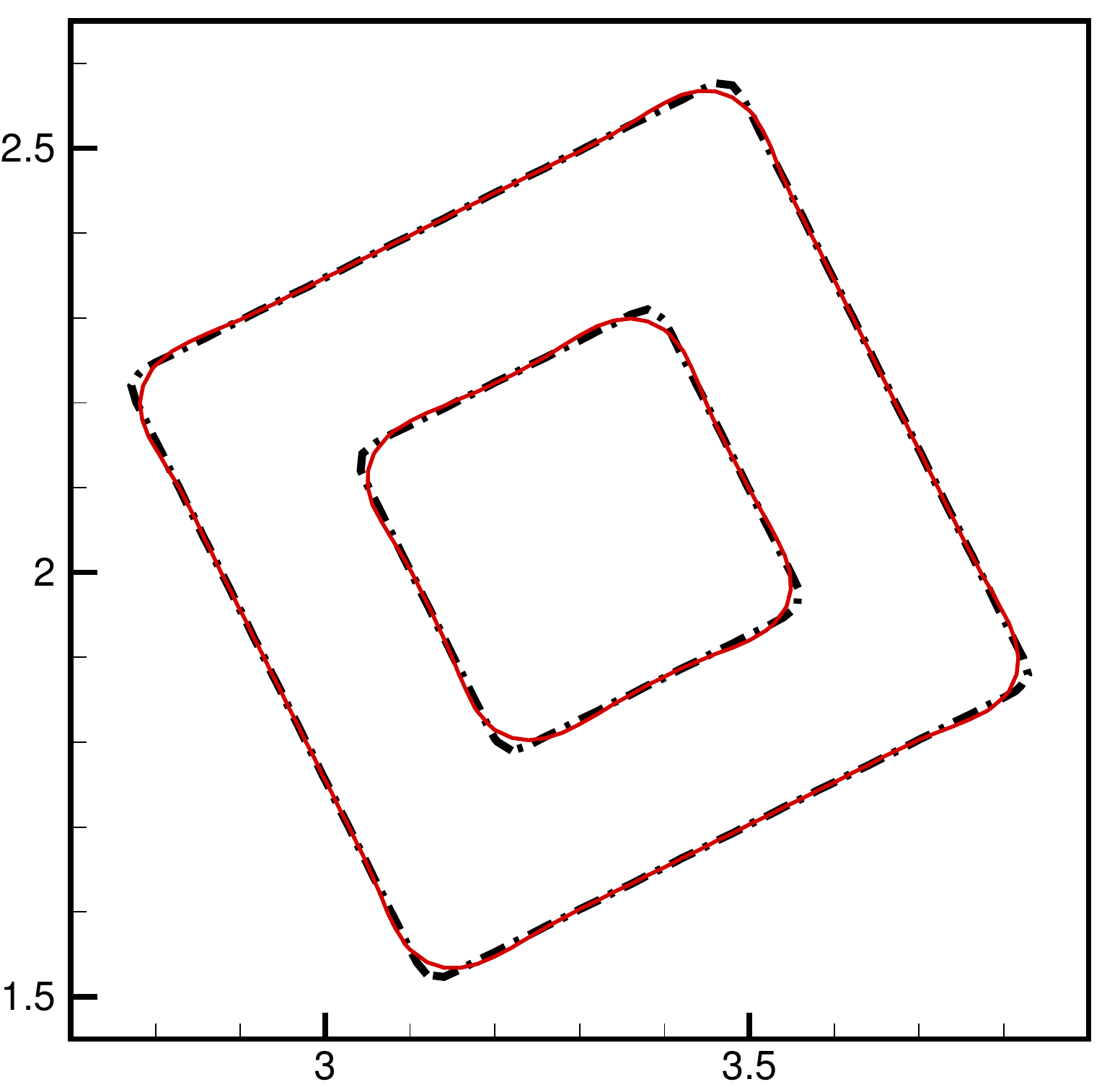} }
	\subfigure[] {
		\centering
		\includegraphics[width=0.3\textwidth]{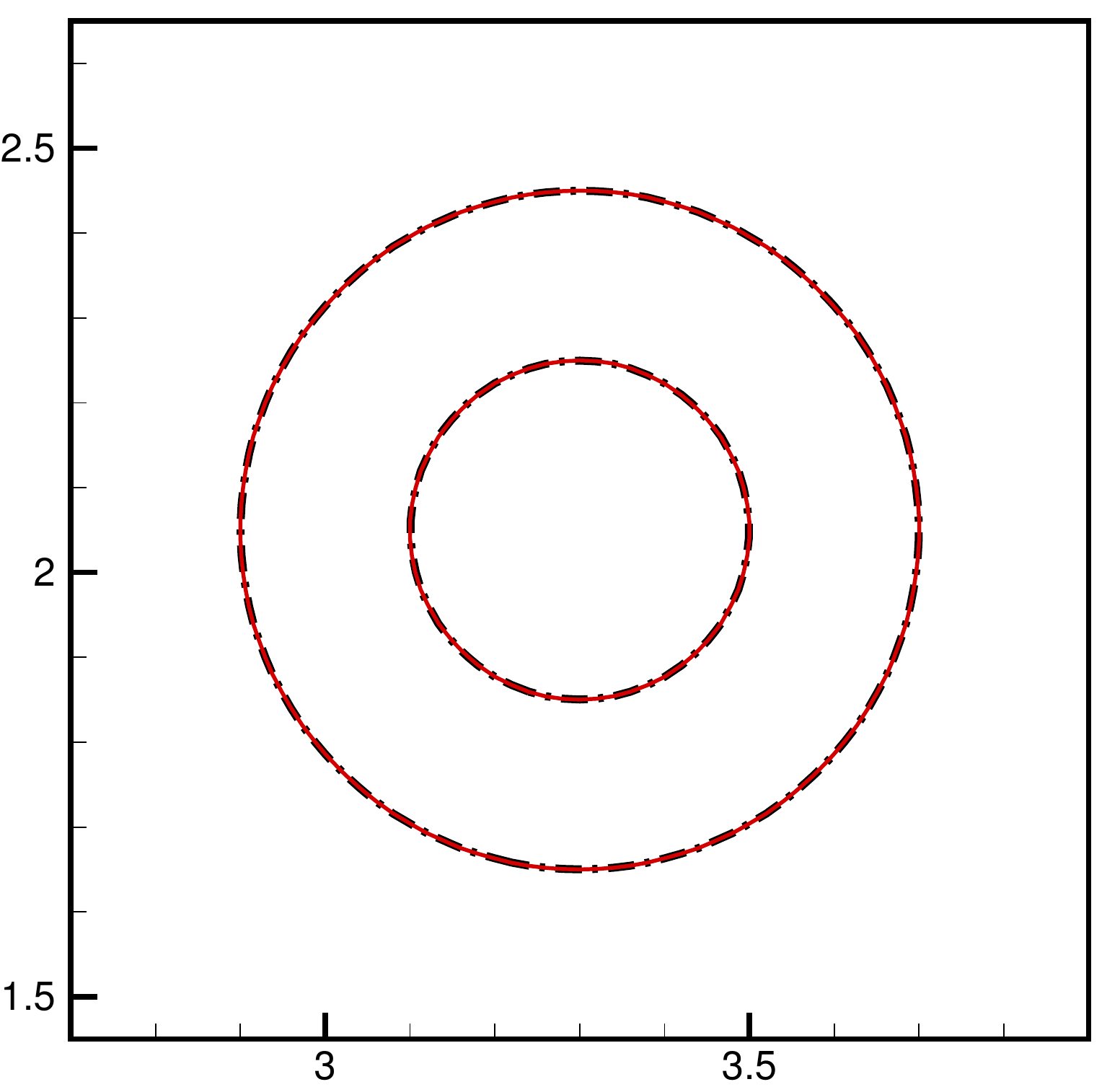} }
	\subfigure[] {
		\centering
		\includegraphics[width=0.3\textwidth]{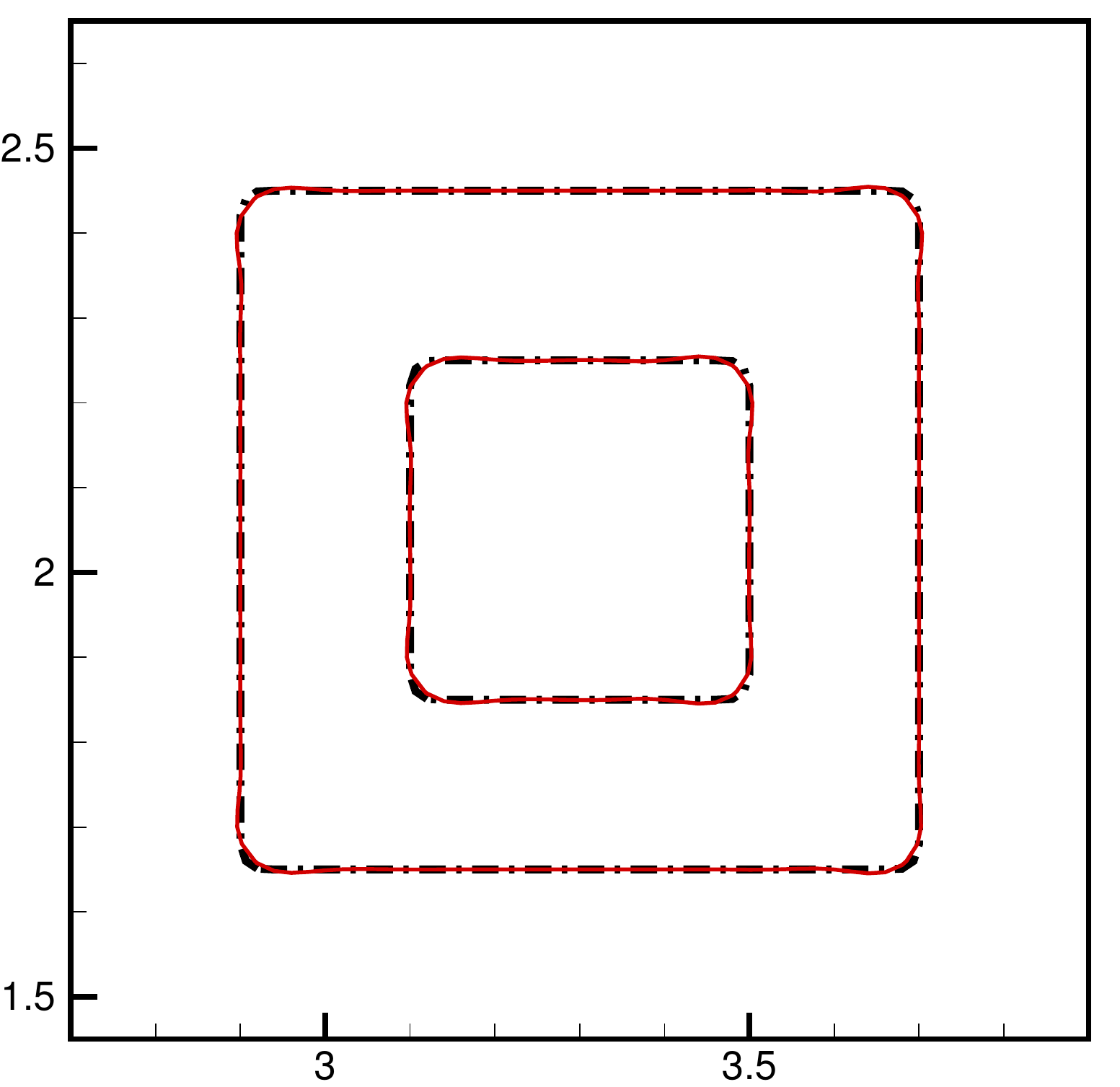} }
	\subfigure[] {
		\centering
		\includegraphics[width=0.3\textwidth]{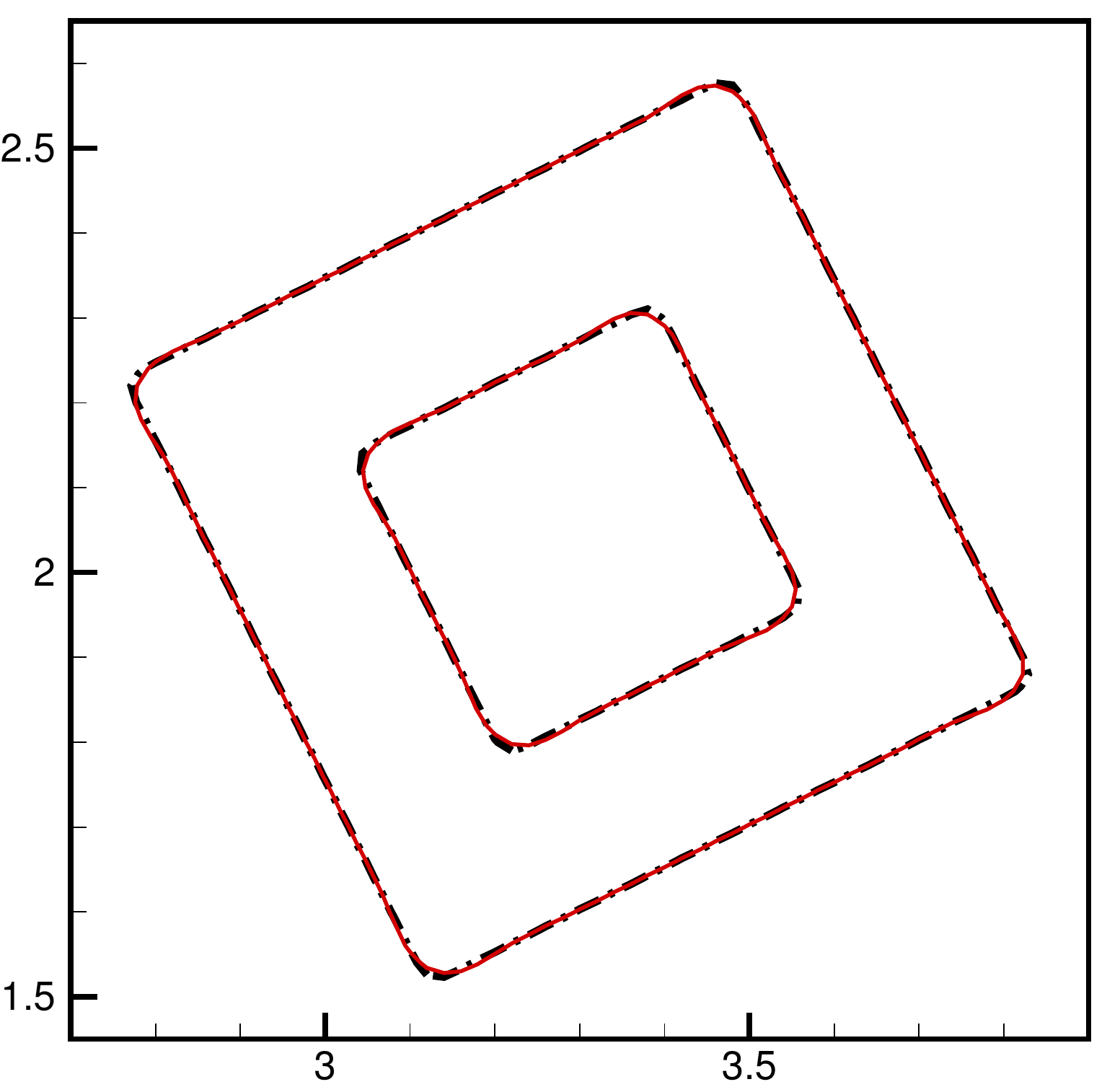} }
	\subfigure[] {
		\centering
		\includegraphics[width=0.3\textwidth]{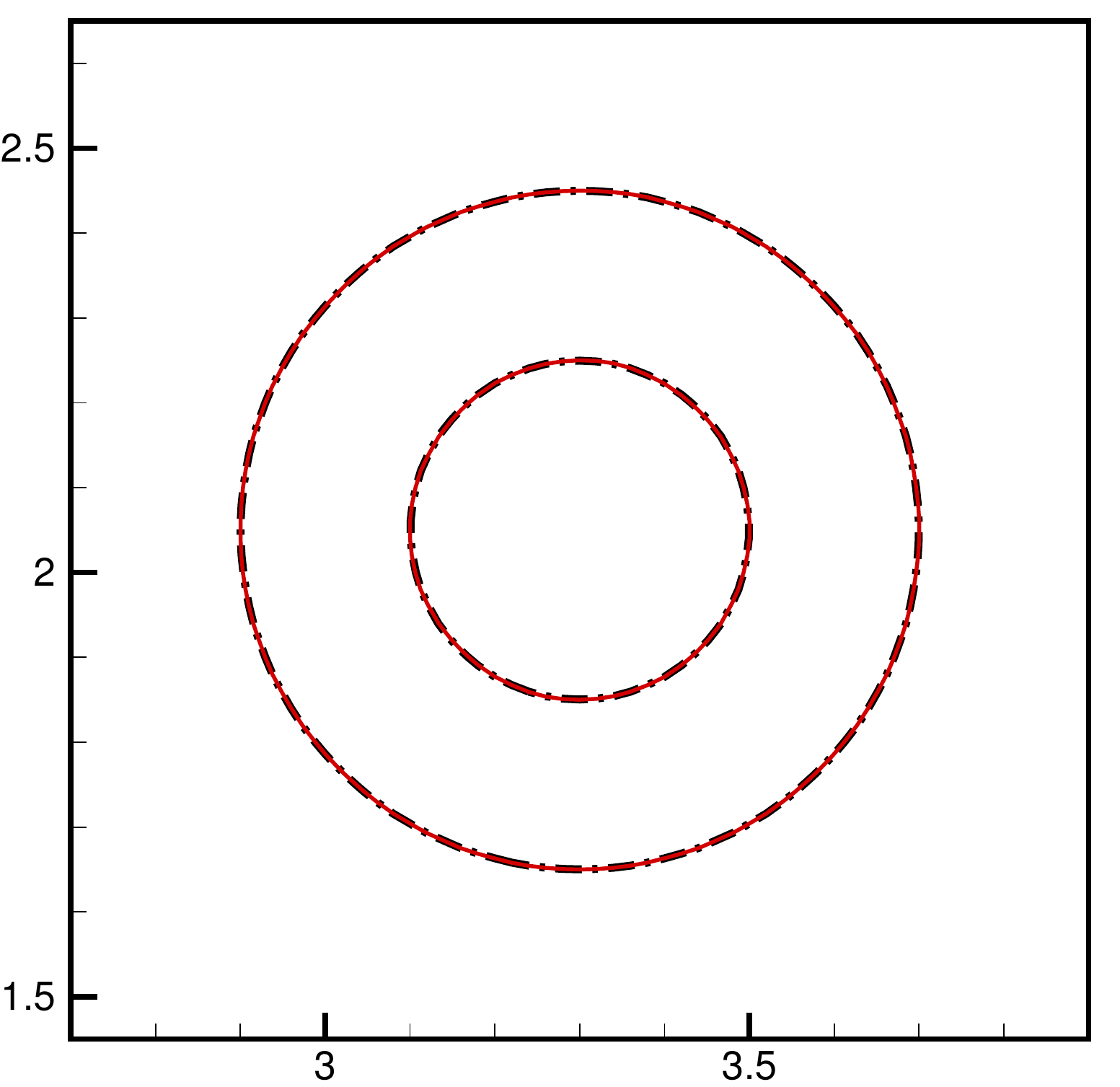} }
	\caption{Numerical results of the Rudman translation test using the second-order (a-c) and fourth-order (d-f) surface polynomials on a $200 \times 200$ mesh. Black dash dot line is the exact solution and red solid line is the numerical solution. Both lines indicate 0.5-contour of the VOF field.}
	\label{Tran-Rudman-Figure}
\end{figure}

We firstly conduct the translation test proposed by Rudman \cite{Rudman1997}. The computational domain $[0, 4]\times[0,4]$ is divided into $200 \times 200$ cells. Three types of interfaces are separately translated by a constant velocity field (2, 1), which are (a) a hollow square, (b) a tilted ($26.57^\circ$) hollow square and (c) a hollow circle. The length or radius of the outer and inner interfaces are 0.8 and 0.4 respectively. Both interfaces are initially centered at (0.8, 0.8). A CFL number of 0.25 is used in all the computations. As shown in \autoref{Tran-Rudman-Figure}, the numerical results of both surface polynomials ($p=2,4$) coincide well with the exact solution. The major error of the THINC/LS method comes from the corners of the square, where the right angle is rounded to some extent during the computation. Compared to the 2nd-order surface polynomial, the 4th-order surface polynomial can keep the better-resolved sharp conner. 

Quantitative comparisons of the THINC/LS method with other existing VOF methods are given in \autoref{Tran-Rudman-Table}. It is observed that the THINC/LS method can produce more accurate results than the VOF methods especially in the hollow-circle case  where high-order surface representation is preferred.

\begin{table}[htbp]
	\begin{center}
    \protect\caption{Numerical errors ($E_r$) of different methods in the Rudman translation test $(\beta=3.5/\Delta x)$} 
	\label{Tran-Rudman-Table}
		\begin{tabular}{lccc}
		\hline
Methods &	Hollow square ($E_r$) &	Hollow titled square ($E_r$) &	Hollow circle ($E_r$) \\
		\hline
SLIC \cite{Rudman1997} &	$1.32\times10^{-1}$ &	$1.08\times10^{-1}$ &	$9.18\times10^{-2}$ \\
Hirt-Nichols \cite{Rudman1997} &	$6.86\times10^{-3}$ &	$1.60\times10^{-1}$ &	$1.90\times10^{-1}$ \\
FCT-VOF \cite{Rudman1997} &	$1.63\times10^{-8}$ &	$8.15\times10^{-2}$ &	$3.99\times10^{-2}$ \\
Youngs \cite{Rudman1997} &	$2.58\times10^{-2}$ &	$3.16\times10^{-2}$ &	$2.98\times10^{-2}$ \\
Stream/Youngs \cite{Harvie2000} &	$2.70\times10^{-2}$ &	$3.08\times10^{-2}$ &	$2.66\times10^{-2}$ \\
Stream/Puckett\cite{Harvie2000} &	$3.33\times10^{-2}$ &	$3.15\times10^{-2}$ &	$6.96\times10^{-3}$ \\
		\hline
THINC/LS (p=2) &	$2.16\times10^{-2}$ &	$2.24\times10^{-2}$ &	$2.13\times10^{-3}$ \\
THINC/LS (p=4) &	$1.34\times10^{-2}$ &	$1.22\times10^{-2}$ &	$2.58\times10^{-3}$ \\
	       \hline       	        
		\end{tabular}
	\end{center}
\end{table}

\subsection{Solid body rotation test}
The Zalesak slotted disk test has been widely used to evaluate interface capturing schemes. As described in \cite{Zalesak}, a slotted circle centered at (0.5, 0.75) with radius 0.15 is rotated in a unit domain. After one revolution period, the slotted circle returns back to its initial state if there not any numerical errors. The slot is defined by ($|x-0.5|\le 0.025$ and $y\le 0.85$) and the velocity field is given by $(0.5-y, x-0.5)$. The CFL number in our computation is about 0.25. 

\begin{figure}[htbp]
	\centering
	\subfigure[] {
		\centering
		\includegraphics[width=0.45\textwidth]{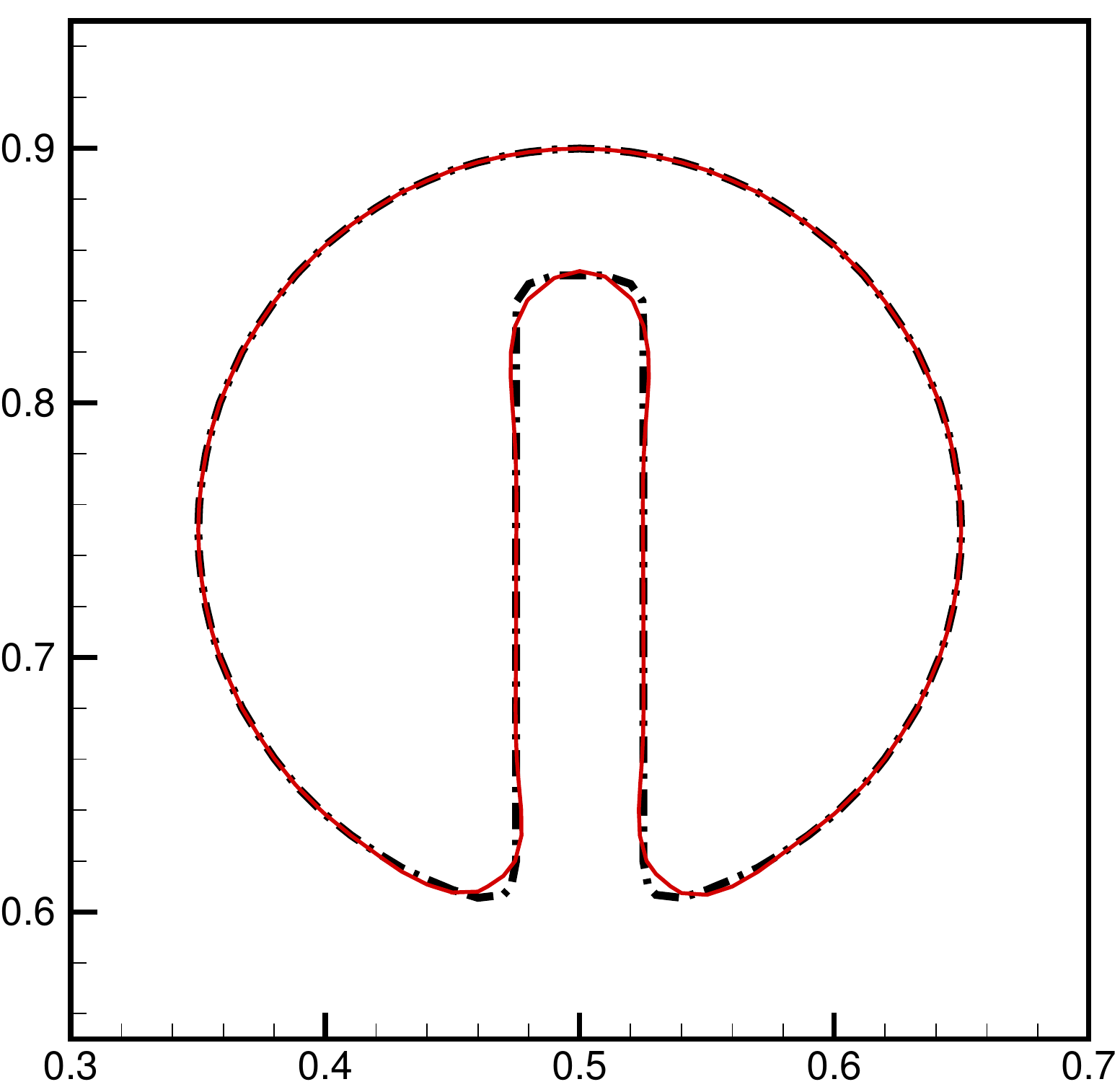} }
	\subfigure[] {
		\centering
		\includegraphics[width=0.45\textwidth]{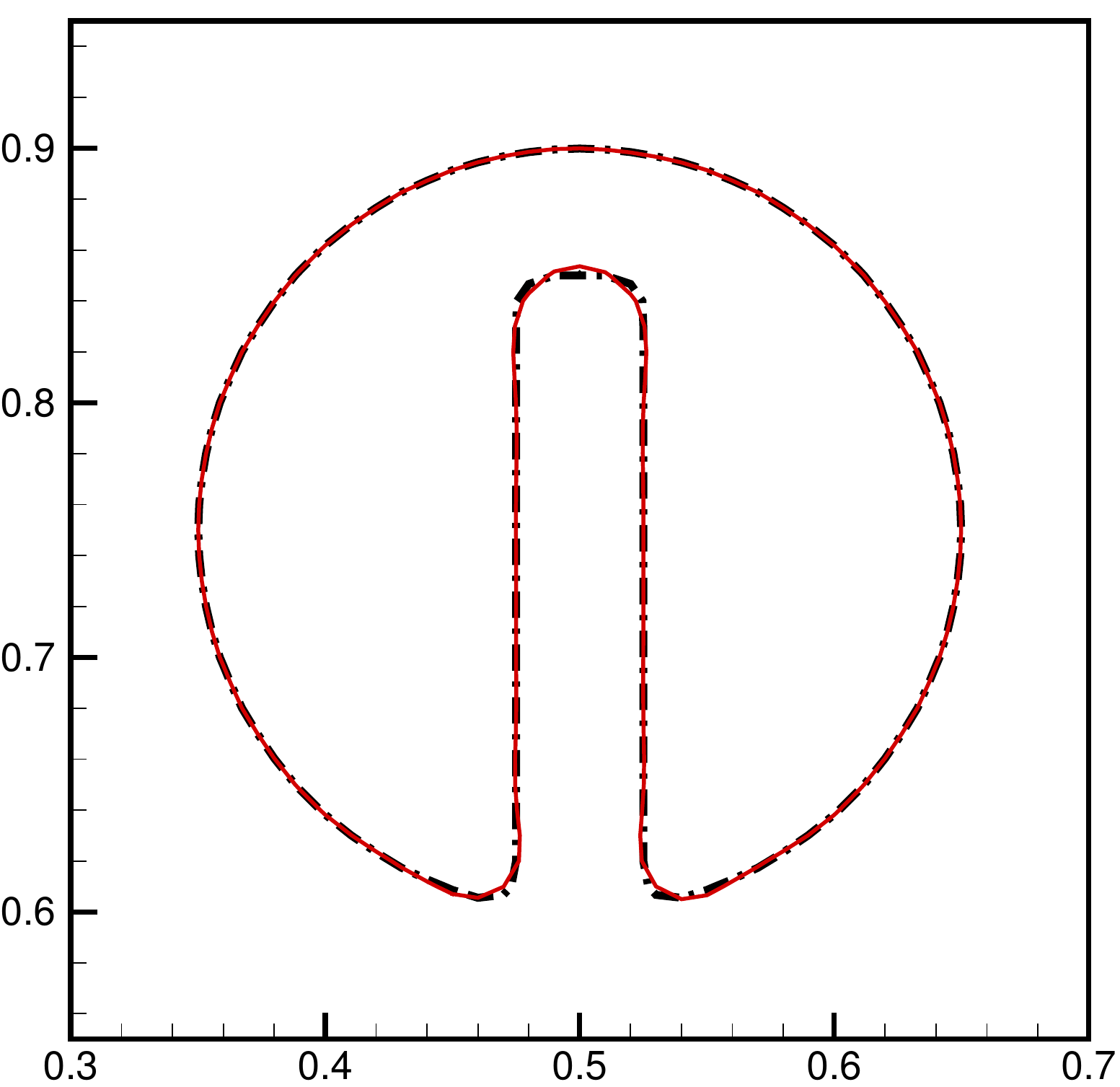} }
	\caption{Numerical results of the Zalesak rotation test after one revolution using (a) second-order and (b) fourth-order (b) surface polynomials on a $100 \times 100$ mesh. Black dash dot line is the exact solution and red solid line is the numerical solution. Both lines are the 0.5-contours of the VOF fields.}
	\label{Rot-Zalesak-Figure}
\end{figure}

\begin{figure}[htbp]
	\centering
	\subfigure[] {
		\centering
		\includegraphics[width=0.45\textwidth]{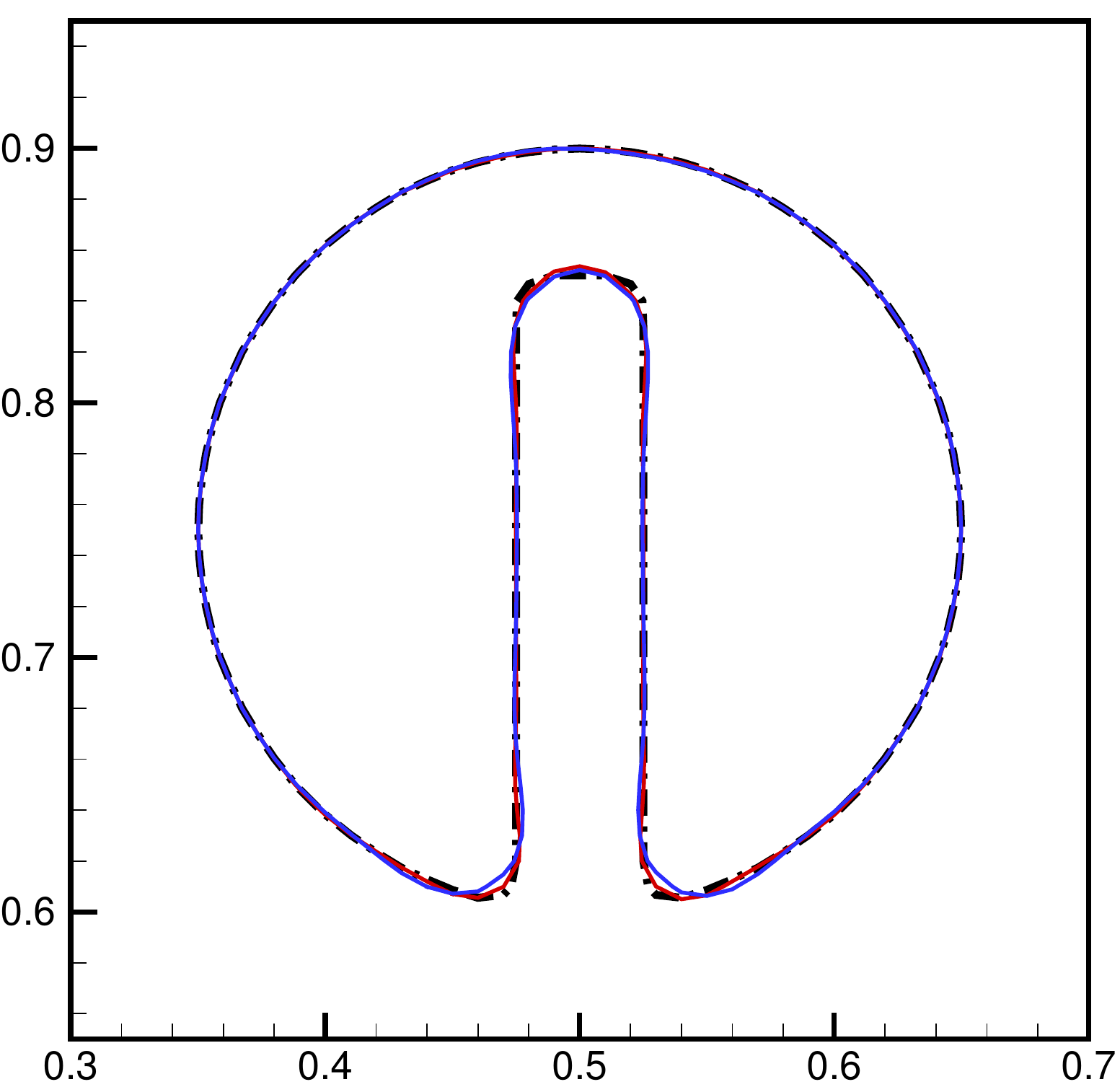} } 
	\subfigure[] {
		\centering
		\includegraphics[width=0.45\textwidth]{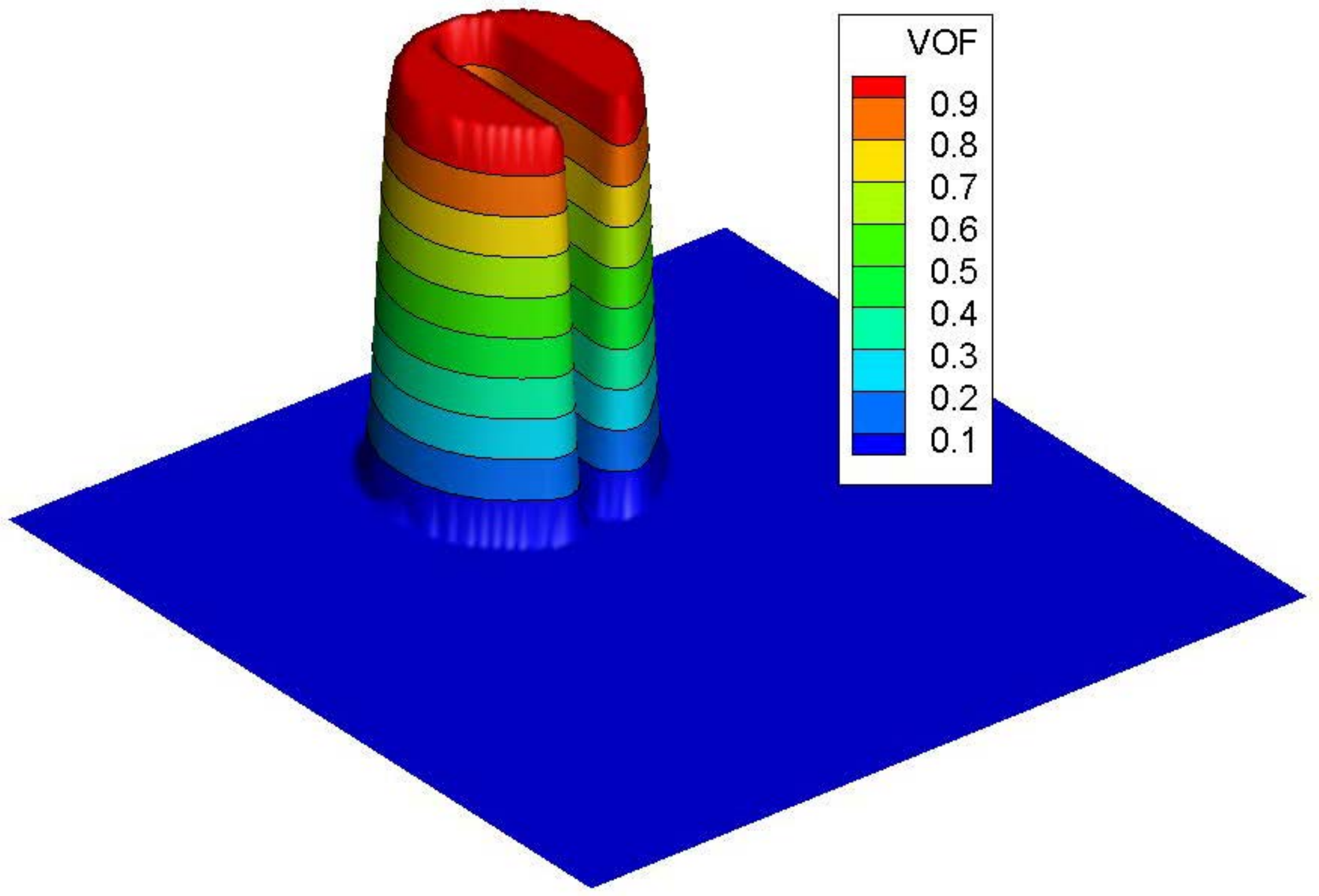} }
	\caption{The 0.5-contour and bird's eye view of the VOF function in the Zalesak rotation test after ten revolution periods using fourth-order surface polynomial on a $100 \times 100$ mesh. In (a), black dash dot line stands for the exact solution, red and blue solid lines are the numerical solutions after one and ten revolutions respectively.}
	\label{Rot-Zalesak-Figure-ten}
\end{figure}

Numerical results of THINC/LS method with different orders of the surface polynomials are illustrated in \autoref{Rot-Zalesak-Figure}. As Lopez \cite{Lopez2004} pointed out, the interface reconstruction errors are mainly concentrated at the regions of large interface curvatures, such as the corners of the slot. In most of the VOF methods, the slot of the circle is always found to be distorted after one revolution. However, we don't observe significant distortions of the slot in the results of THINC/LS method that uses high-order interface representations. Moreover, the slotted disk still keeps very well after ten revolutions as shown in \autoref{Rot-Zalesak-Figure-ten}. It reveals that the THINC/LS method is very geometrically faithful and effective in preserving the shape of the transported interface.

We give the numerical errors after one revolution and compare it with other THINC schemes in \autoref{Rot-Zalesak-Table1}, the THINC/LS method produces superior results than other THINC schemes. 

In order to evaluate the computational cost, show the elapse time of the THINC/LS method in comparison with the level set method in \autoref{Rot-Zalesak-Table2}. The  THINC/LS method takes nearly twice computational time compared to the level set method, but still within the acceptable range.

\begin{table}[htbp]
	\begin{center}
    \protect\caption{Numerical errors ($E_r$) and convergence rates of Zalesak rotation test after one revolution period $(\beta=3.5/\Delta x)$} 
    \label{Rot-Zalesak-Table1}
		\begin{tabular}{lccccc}
		\hline
Methods &	$50^2$ &	Order &	$100^2$ &	Order &	$200^2$ \\
		\hline
MTHINC \cite{MTHINC} &	$2.93\times10^{-2}$ &	0.86 &	$1.61\times10^{-2}$ &	1.03 &	$7.91\times10^{-3}$  \\
UMTHINC \cite{THINCQQ} &	$8.12\times10^{-2}$ &	1.63 &	$2.61\times10^{-2}$ &	0.97 &	$1.33\times10^{-2}$  \\
THINC/QQ \cite{THINCQQ} &	$8.96\times10^{-2}$ &	1.47 &	$3.22\times10^{-2}$ &	0.95 &	$1.67\times10^{-2}$  \\
	       \hline 
THINC/LS (p=2) &	$1.23\times10^{-1}$ &	3.04 &	$1.50\times10^{-2}$ &	1.07 &	$7.13\times10^{-3}$ \\
THINC/LS (p=4) &	$4.83\times10^{-2}$ &	2.34 &	$9.55\times10^{-3}$ &	1.26 &	$4.00\times10^{-3}$ \\
	       \hline       	        
		\end{tabular}
	\end{center}
\end{table}

\begin{table}[htbp]
	\begin{center}
    \protect\caption{Computation costs of different methods in Zalesak rotation test $(\beta=3.5/\Delta x)$}
    \label{Rot-Zalesak-Table2}
  		\begin{tabular}{lccccc}
  	       \hline  		
Methods &	One revolution period &	Ten revolution period \\
	       \hline  
LS (Narrow band) & 12.15s & 125.36s \\
	       \hline  
THINC/LS (p=2) & 23.01s & 237.98s \\
THINC/LS (p=4)	&  36.27s & 370.11s  \\
	       \hline  
		\end{tabular}
	\end{center}
\end{table}

\begin{figure}[htbp]
	\centering
	\subfigure[] {
		\centering
		\includegraphics[width=0.45\textwidth]{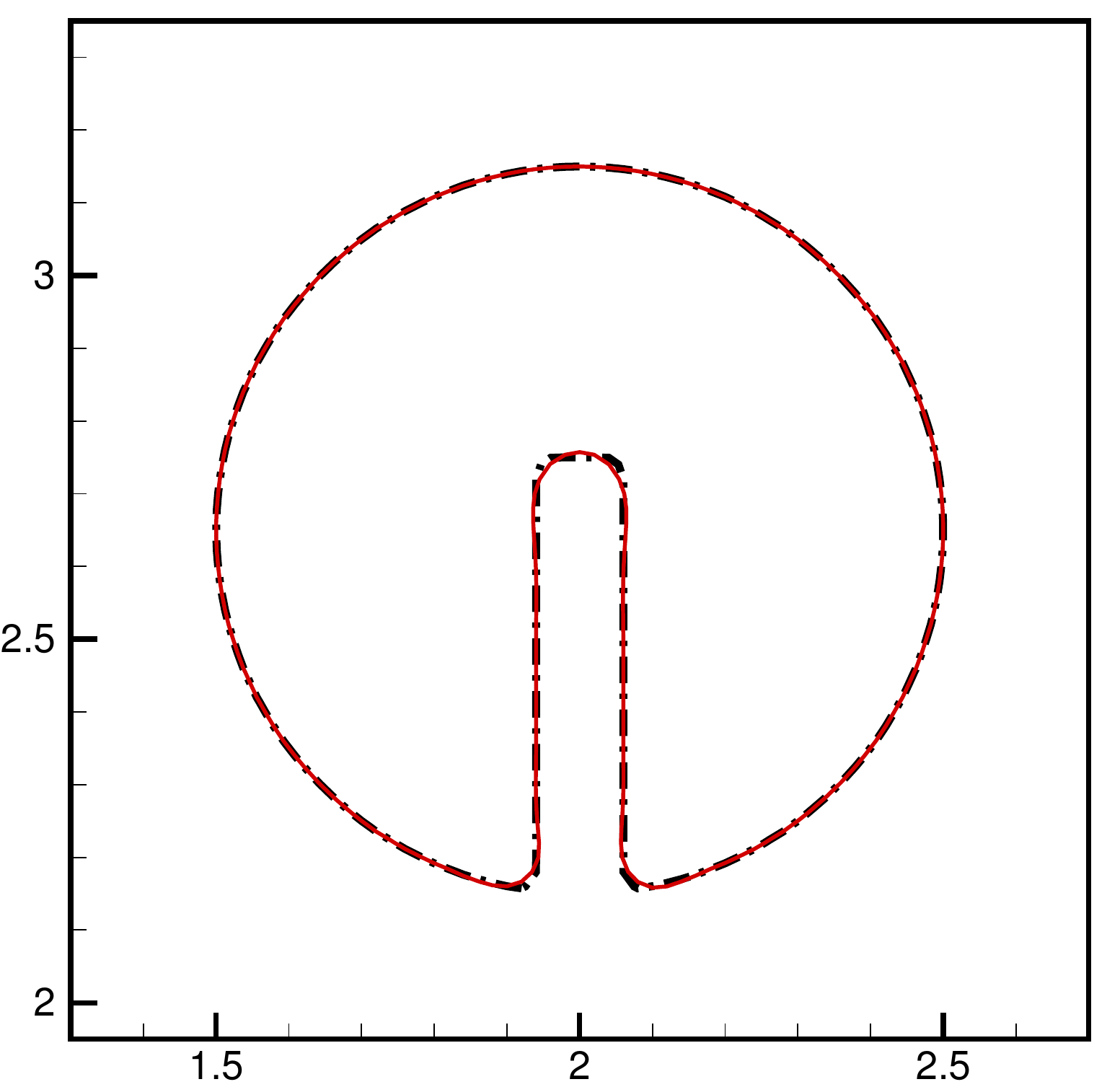} }
	\subfigure[] {
		\centering
		\includegraphics[width=0.45\textwidth]{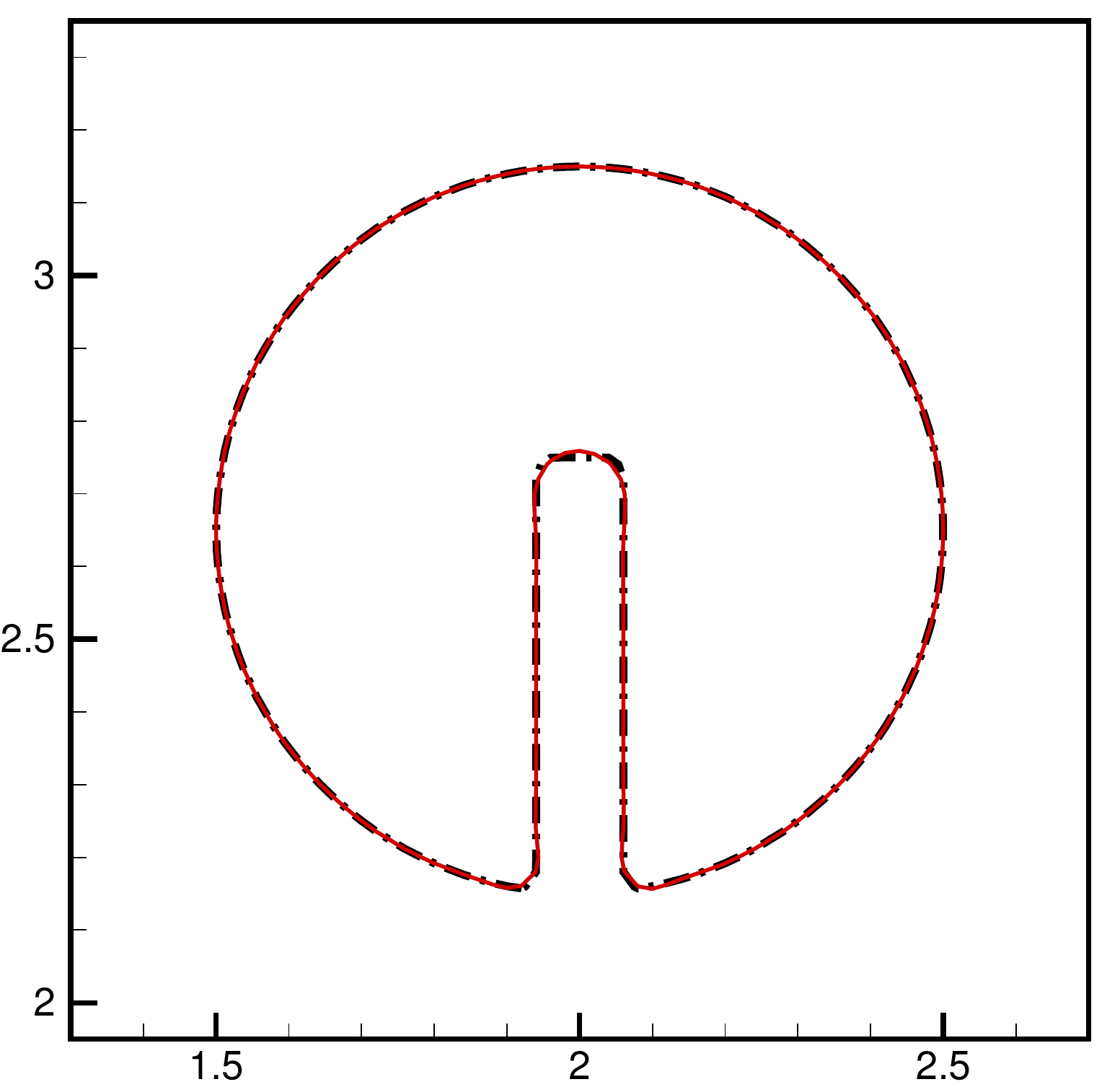} }
	\caption{Numerical results of the Rudman rotation test after one revolution using (a) second-order and (b) fourth-order (b) surface polynomials on a $200 \times 200$ mesh. Black dash dot line is the exact solution and red solid line is the numerical solution. Both lines are the 0.5-contours of the VOF fields.}
	\label{Rot-Rudman-Figure}
\end{figure}

We also conducted the solid rotation benchmark test proposed by Rudman. As described in \cite{Rudman1997}, a slotted circle with radius 0.5 is centered at (2.0, 2.65) on a $[0, 4]\times[0,4]$ domain. The slot is defined by ($|x-2.0|\le 0.06$ and $y\le 2.75$) and the velocity field is given by ($2.0-y, x-2.0$). The computations were conducted on a $200 \times 200$ mesh and the maximum CFL number is about 0.25. \autoref{Rot-Rudman-Figure} shows the numerical results of the THINC/LS method with different order surface polynomials. As discussed before, the main error of the THINC/LS method is concentrated at the slot corners. Nevertheless, the slotted circle in general coincides well with the exact solution without significant distortions. To illustrate this, we further compute the numerical errors of the THINC/LS method and compare with other VOF methods. Shown in \autoref{Rot-Rudman-Table}, the results of THINC/LS method are among the most accurate ones.

\begin{table}[htbp]
	\begin{center}
    \protect\caption{Numerical errors ($E_r$) of the Rudman rotation test $(\beta=3.5/\Delta x)$} 
    \label{Rot-Rudman-Table}
		\begin{tabular}{lc}
		\hline
Methods &	Error \\
	       \hline 
Hirt-Nichols \cite{Hirt1981} &	$9.62\times10^{-2}$ \\
SLIC \cite{SLIC1976} &	$8.38\times10^{-2}$ \\
FCT-VOF \cite{Rudman1997} &	$3.29\times10^{-2}$ \\
Youngs \cite{Rudman1997} &	$1.09\times10^{-2}$ \\
Stream/Youngs \cite{Harvie2000} &	$1.07\times10^{-2}$ \\
Stream/Puckett \cite{Harvie2000} &	$1.00\times10^{-2}$ \\
DDR/Youngs \cite{Harvie2001} &	$1.56\times10^{-2}$ \\
DDR/Puckett \cite{Harvie2001} &	$1.50\times10^{-2}$ \\
EMFPA/Youngs \cite{Lopez2004} &	$1.06\times10^{-2}$ \\
EMFPA/Puckett \cite{Lopez2004} &	$9.73\times10^{-3}$ \\
EMFPA/SIR \cite{Lopez2004} &	$8.74\times10^{-3}$ \\
THINC \cite{THINC} &	$3.52\times10^{-2}$ \\
THINC/WLIC \cite{THINCWLIC} &	$1.96\times10^{-2}$ \\
THINC/SW \cite{THINCSW} &	$1.34\times10^{-2}$ \\
THINC/QQ \cite{THINCQQ} &	$1.42\times10^{-2}$ \\
Linear fit/Largragian \cite{Scardovelli2003} &	$9.42\times10^{-3}$ \\
Quadratic fit/Lagrangian \cite{Scardovelli2003} &	$5.47\times10^{-3}$ \\
Quadratic fit+continuity/Lagrangian \cite{Scardovelli2003} &	$4.16\times10^{-3}$ \\
	       \hline 
THINC/LS (p=2) & 	$5.09\times10^{-3}$ \\
THINC/LS (p=4) &	$3.77\times10^{-3}$ \\
	       \hline       	        
		\end{tabular}
	\end{center}
\end{table}

\subsection{Vortex deformation transport test}
We further assessed the THINC/LS method with the moving interface advected by  a deformational velocity field. As described in \cite{Rider1998}, a circle with radius 0.15 is initially centred at (0.5, 0.75) in a unit domain. A time-dependent velocity field is given by the stream function
\begin{equation}
\Psi(x,y,t)=\frac{1}{\pi}sin^2(\pi x)sin^2(\pi y)cos(\frac{\pi t}{T}).
\label{deform-vel}
\end{equation}

Advected by velocity field \eqref{deform-vel}, the initial circle is firstly distorted and stretched into a spiral shape with a thin tail up to $t=T/2$, and then transported at a reverse velocity field after $T/2$ until restoring its initial shape at $T$. For a large $T$, the width of the tail can be thiner the size of mesh cell, and thus not resolvable by the Eulerian fixed grid at $T/2$. 
It is well known that the conventional VOF methods tend to generate small droplets or  "flotsams" at $t = T/2$ when the tail of the spiral-like VOF field is stretched into a thin film under grid resolution. In this case, the interface losses its geometrical information, and it is hard to restore the initial state at $t = T$.

\begin{figure}[htbp]
	\centering
	\subfigure[] {
		\centering
		\includegraphics[width=0.45\textwidth]{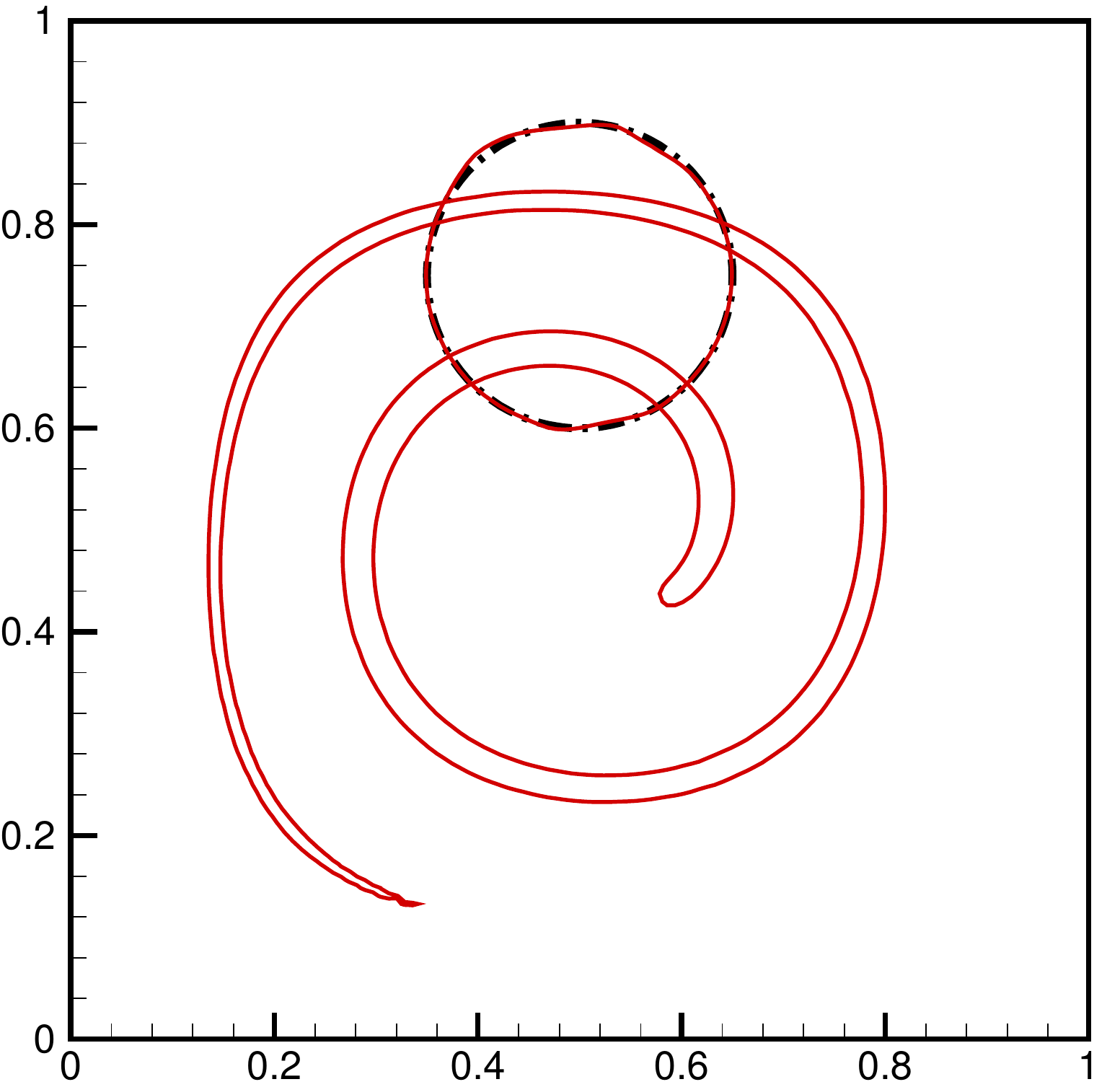} }
	\subfigure[] {
		\centering
		\includegraphics[width=0.45\textwidth]{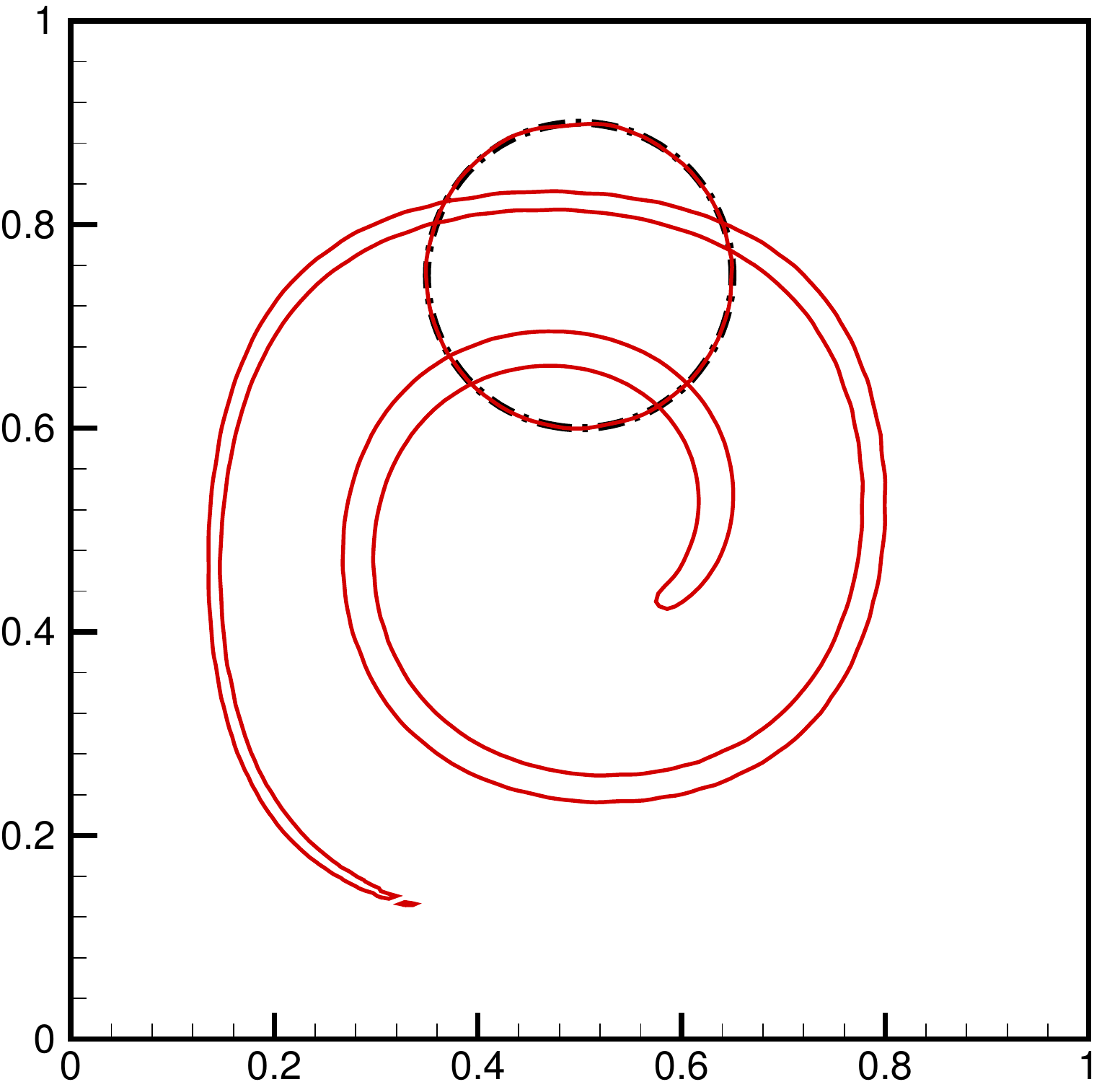} }
	\caption{Numerical results of the Rider-Kothe single vortex test using (a) second-order and (b) fourth-order surface polynomial on a $128 \times 128$ mesh. Solid red line is the numerical results at $t=T/2,T$ and dashed black line is the exact solution. Both lines are the  the 0.5-contours of the VOF fields.}
	\label{Shr-Rider-Figure}
\end{figure}

\begin{figure}[htbp]
	\centering
	\subfigure[] {
		\centering
		\includegraphics[width=0.45\textwidth]{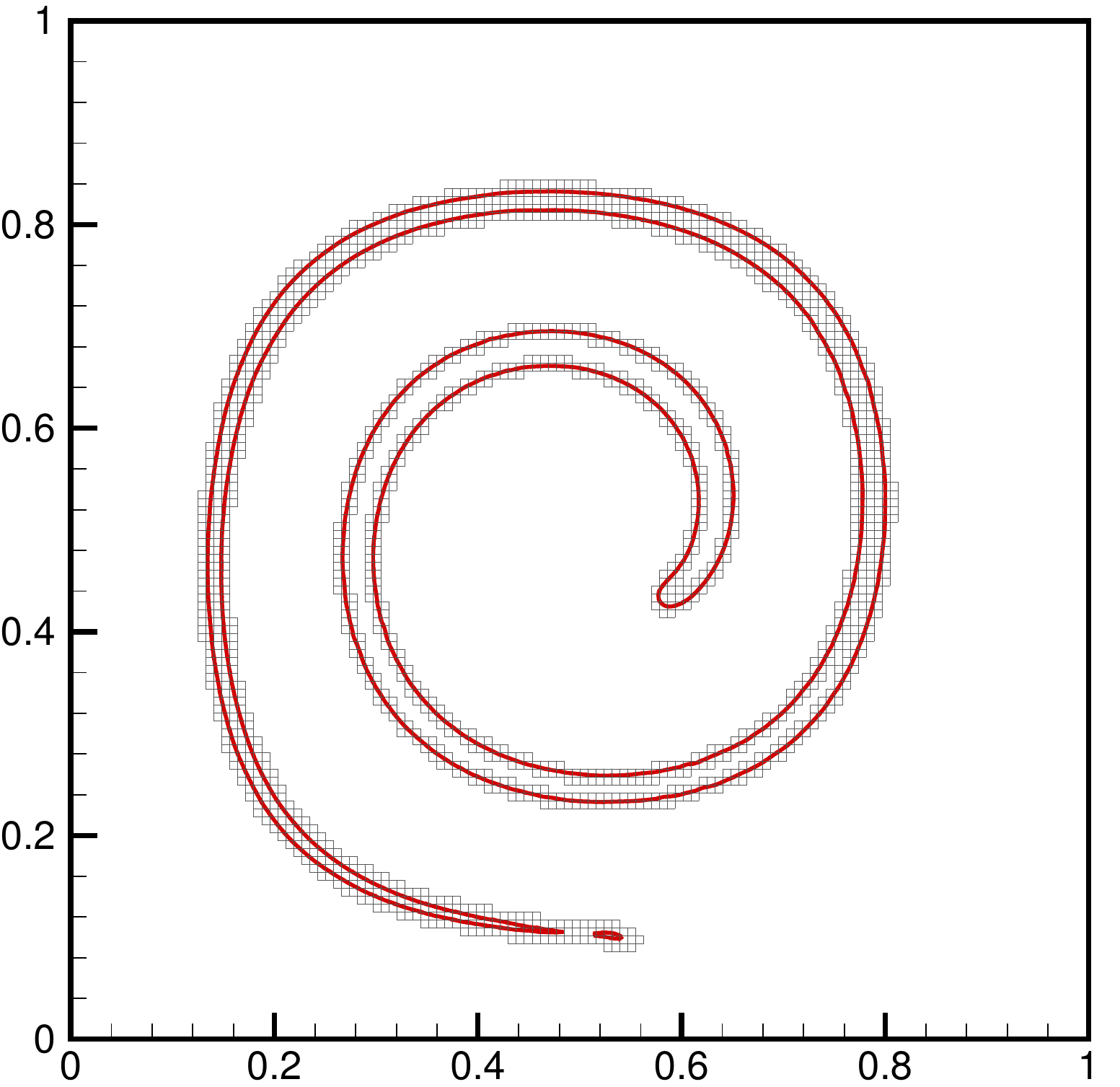} }
	\subfigure[] {
		\centering
		\includegraphics[width=0.45\textwidth]{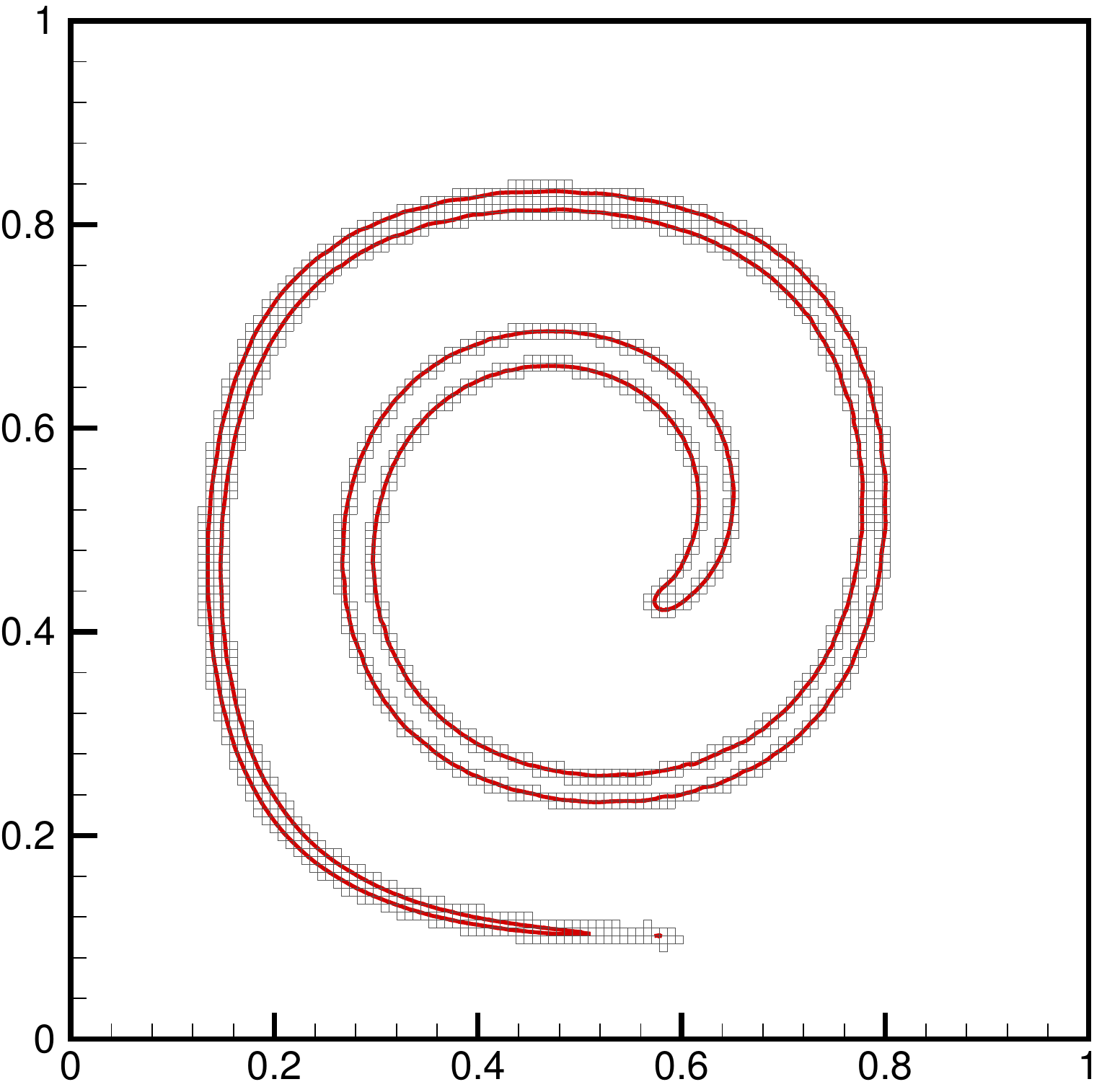} }
	\caption{Reconstructed interfaces of the Rider-Kothe single vortex test using (a) second-order and (b) fourth-order surface polynomials on a $128 \times 128$ mesh.}
	\label{Shr-Rider-Figure-Interface}
\end{figure}

We tested the case of $T=8$ on grids with gradually increased resolutions, $32 \times 32$, $64 \times 64$, $128 \times 128$. The maximum CFL number of all computations are set to 0.25. The numerical results on  $128 \times 128$ grid at $t=T/2$ or $t=T$  are shown in \autoref{Shr-Rider-Figure}, where the interface is identified by the 0.5-contour of the VOF values as the center line of the transition layer. It it observed that THINC/LS method with both 2nd and 4th-order surface polynomials can restore the initial circular shape with adequate accuracy, and the 4th-order polynomial give better solution quality. Without explicit geometrical reconstruction, THINC algorithm exactly conserves the transported field but allows a finite thickness (usually 2 or 3 cell-widths) for the interface transition layer. So, the cells where the VOF values are smaller than 0.5 are not visible by examining only the 0.5 contour. Due to the velocity field  \eqref{deform-vel}, the spiral tail is heavily stretched in the flow direction at $T/2$, which generates VOF values smaller than 0.5 in the surrounding cells. The interface segments in these cell can be identified by \eqref{eq-interface}. We plot the reconstructed interface at $t=T/2$ using \eqref{eq-interface} in \autoref{Shr-Rider-Figure-Interface}. It reveals that the THINC/LS method can reproduce  the interface even when the interface is under the grid resolution.  Unlike some complex VOF schemes using two non-contiguous linear segments \cite{Lopez2005} or markers \cite{Aulisa2003} to resolve the thin interfaces, the THINC/LS method is much  easer to implement in 3D and unstructured grids. We further show the $L_1$ VOF error and convergence rate of the THINC/LS method in comparison with other VOF and hybrid VOF methods in \autoref{Shr-Rider-Table}. It is observed that the THINC/LS method can produce competitive results compared to other hybrid VOF methods.

\begin{table}[htbp]
	\begin{center}
    \protect\caption{Numerical errors and convergence rates of the time-dependent reversing single vortex test ($\beta=7.0/\Delta x$)} 
	\label{Shr-Rider-Table}
		\begin{tabular}{lccccc}
		\hline
Methods &	$32^2$ &	Order &	$64^2$ &	Order &	$128^2$ \\
	       \hline 
Rider-Kother/Puckett \cite{Rider1998} &	$4.78\times10^{-2}$ &	2.78 &	$6.96\times10^{-3}$ &	2.27 &	$1.44\times10^{-3}$ \\
Stream/Puckett \cite{Harvie2000} &	$3.72\times10^{-2}$ &	2.45 &	$6.79\times10^{-3}$ &	2.52 &	$1.18\times10^{-3}$ \\
Stream/Youngs \cite{Harvie2000} &	$3.61\times10^{-2}$ &	1.85 &	$1.00\times10^{-2}$ &	2.21 &	$2.16\times10^{-3}$ \\
EMFPA/Puckett \cite{Lopez2004} &	$3.77\times10^{-2}$ &	2.52 &	$6.58\times10^{-3}$ &	2.62 &	$1.07\times10^{-3}$ \\
THINC/WLIC \cite{THINCWLIC} &	$4.16\times10^{-2}$ &	1.37 &	$1.61\times10^{-2}$ &	2.18 &	$3.56\times10^{-3}$ \\
THINC/SW \cite{THINCSW} &	$3.90\times10^{-2}$ &	1.36 &	$1.52\times10^{-2}$ &	1.94 &	$3.96\times10^{-3}$ \\
THINC/QQ \cite{THINCQQ} &	$6.70\times10^{-2}$ &	2.14 &	$1.52\times10^{-2}$ &	2.31 &	$3.06\times10^{-3}$ \\     
Hybrid markers-VOF \cite{Aulisa2003} &	$2.53\times10^{-2}$ &	3.19 &	$2.78\times10^{-3}$ &	2.54 &	$4.78\times10^{-4}$ \\
Markers-VOF \cite{Lopez2005} &	$7.41\times10^{-3}$ &	1.83 &	$2.12\times10^{-3}$ &	2.31 &	$4.27\times10^{-4}$ \\
	       \hline        	       	       
THINC/LS (p=2) &	$1.04\times10^{-1}$ &	2.66 &	$1.65\times10^{-2}$ &	3.39 &	$1.57\times10^{-3}$\\
THINC/LS (p=4) &	$2.92\times10^{-2}$ &	2.85 &	$4.04\times10^{-3}$ &	2.52 &	$7.06\times10^{-4}$\\
	       \hline       	        
		\end{tabular}
	\end{center}
\end{table}

\begin{figure}[htbp]
	\centering
	\subfigure[] {
		\centering
		\includegraphics[width=0.45\textwidth]{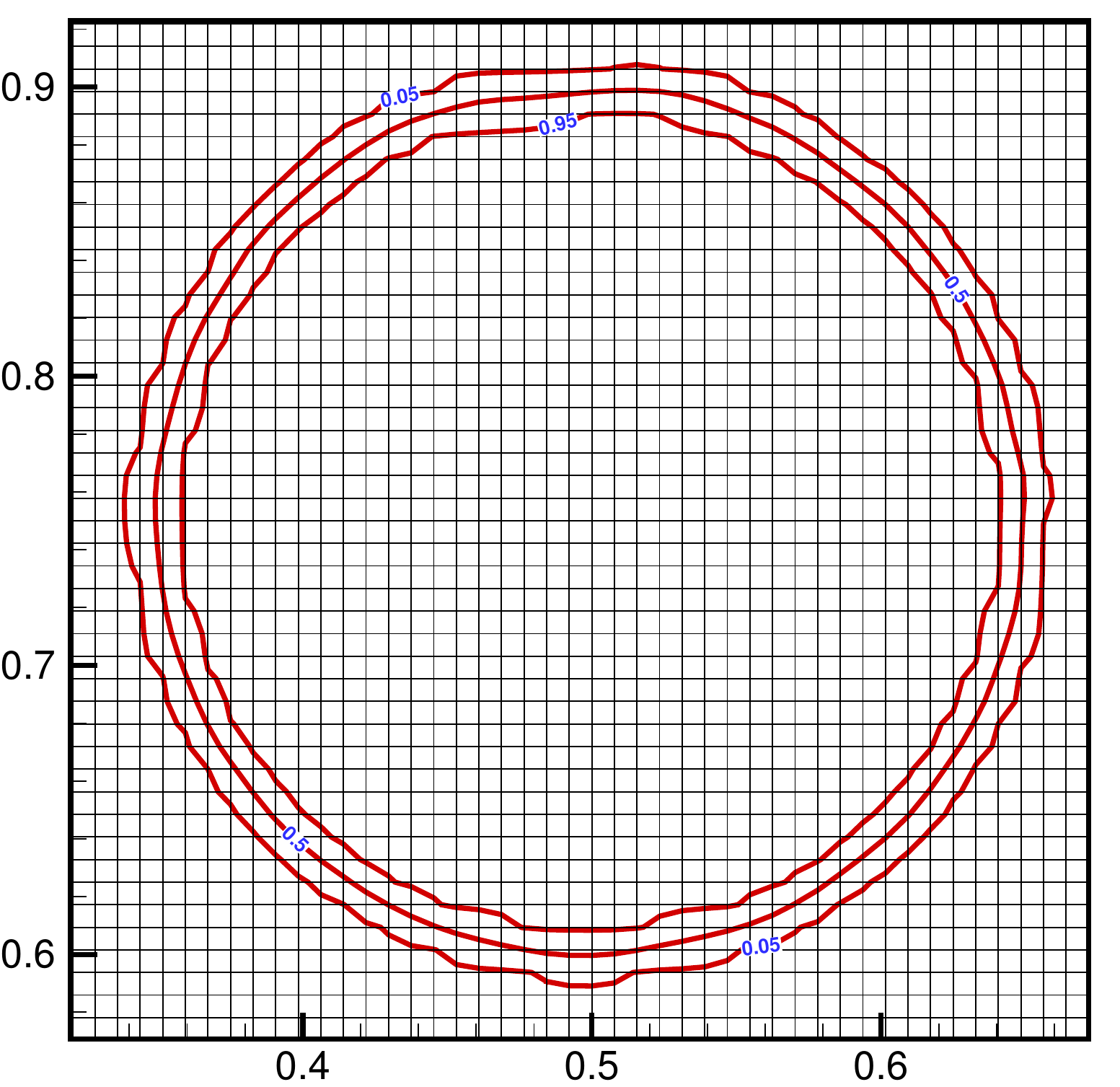} }
	\subfigure[] {
		\centering
		\includegraphics[width=0.45\textwidth]{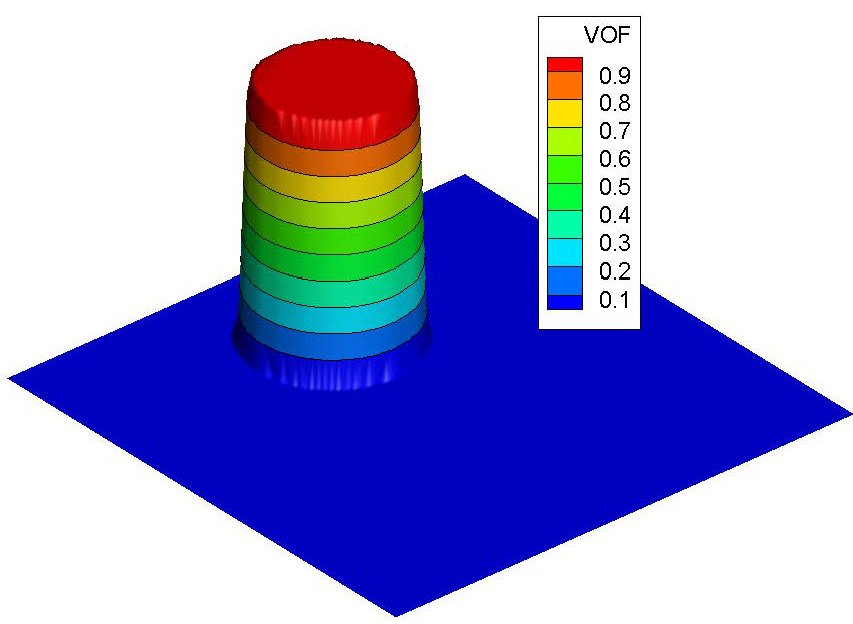} }
	\caption{The 0.05, 0.5 and 0.95 contours and bird's eye view of VOF function in the Rider-Kothe single vortex rotation test after one period ($t=T/2$) using the 4-order surface polynomial on a $128 \times 128$ mesh.}
	\label{Shr-Rider-Diff}
\end{figure}

We also examined the numerical diffusion of the THINC/LS method. Shown in \autoref{Shr-Rider-Diff}, the interface thickness and sharpness maintain satisfactorily  after one period $t=T$.

\section{Conclusions}
We propose a novel interface capturing scheme, so-called THINC/LS, for capturing moving interfaces. The innovative and practically-significant aspects of the proposed THINC/LS schemes are summarized as follows: 
\begin{itemize}
\item Higher-order (arbitrary order in principle) surface polynomials can be determined from the smooth level set function, and then used to build the THINC or VOF function. 
\item The interface in the level set field is synchronized with the interface identified from the THINC reconstruction with the mass conservation constraint. Thus, the numerical conservativeness is ensured. 
\item The numerical scheme can produce accurate  interfaces even for the interfaces under grid resolution.
\item The numerical scheme maintains well the interface thickness and sharpness, as well as the superior geometrical faithfulness of the moving interface. 
\item The THINC/LS method provides an easy-to-code algorithm for implementation and can be straightforwardly applied to 3D and unstructured grids.
\end{itemize}

Numerical verifications have been conducted via interface reconstruction tests and advection benchmark tests for moving interfaces evaluate the THINC/LS method. It is revealed that the method can significantly improve the numerical accuracy in comparison with other existing methods through much simpler algorithmic procedure. 

\section{Acknowledgments}

The author gratefully thanks to Dr. X. Deng, Dr. B. Xie and P. Jin for their fruitful discussions on this paper. This work was supported in part by the Priority Academic Program Development of Jiangsu Higher Education Institutions (PAPD) and the financial support of the project from the Fundamental Research Funds for the Central Universities NP2016204, as well as the fund from JSPS (Japan Society for the Promotion of Science) under Grant Nos. 15H03916 and 17K18838. 

\clearpage{}
\bibliographystyle{model1-num-names}
\bibliography{THINC_LS} 

\begin{thebibliography}{43}
\expandafter\ifx\csname natexlab\endcsname\relax\def\natexlab#1{#1}\fi
\providecommand{\url}[1]{\texttt{#1}}
\providecommand{\href}[2]{#2}
\providecommand{\path}[1]{#1}
\providecommand{\DOIprefix}{doi:}
\providecommand{\ArXivprefix}{arXiv:}
\providecommand{\URLprefix}{URL: }
\providecommand{\Pubmedprefix}{pmid:}
\providecommand{\doi}[1]{\href{http://dx.doi.org/#1}{\path{#1}}}
\providecommand{\Pubmed}[1]{\href{pmid:#1}{\path{#1}}}
\providecommand{\bibinfo}[2]{#2}
\ifx\xfnm\relax \def\xfnm[#1]{\unskip,\space#1}\fi
\bibitem[{Unverdi and Tryggvason(1992)}]{Tryggvason1992}
\bibinfo{author}{S.~O. Unverdi}, \bibinfo{author}{G.~Tryggvason},
\newblock \bibinfo{title}{{A front-tracking method for viscous,
  incompressible,multi-fluid flows}},
\newblock \bibinfo{journal}{Journal of Computational Physics}
  \bibinfo{volume}{100} (\bibinfo{year}{1992}) \bibinfo{pages}{25--37}.
\bibitem[{Tryggvason et~al.(2001)Tryggvason, Bunner, Esmaeeli, and
  et~al}]{Tryggvason2001}
\bibinfo{author}{G.~Tryggvason}, \bibinfo{author}{B.~Bunner},
  \bibinfo{author}{A.~Esmaeeli}, \bibinfo{author}{et~al},
\newblock \bibinfo{title}{{A Front-Tracking Method for the Computations of
  Multiphase Flow}},
\newblock \bibinfo{journal}{Journal of Computational Physics}
  \bibinfo{volume}{169} (\bibinfo{year}{2001}) \bibinfo{pages}{708--759}.
\bibitem[{Rider and Kothe(1995{\natexlab{a}})}]{Rider1995a}
\bibinfo{author}{W.~J. Rider}, \bibinfo{author}{D.~B. Kothe},
\newblock \bibinfo{title}{{A Marker Particle Method for Interface Tracking}},
\newblock in: \bibinfo{booktitle}{6th International Symposium on Computational
  Fluid Dyanamics}, \bibinfo{year}{1995}{\natexlab{a}}, pp.
  \bibinfo{pages}{976--981}.
\bibitem[{Rider and Kothe(1995{\natexlab{b}})}]{Rider1995b}
\bibinfo{author}{W.~J. Rider}, \bibinfo{author}{D.~B. Kothe},
\newblock \bibinfo{title}{{Stretching and Tearing Interface Tracking Methods}},
\newblock in: \bibinfo{booktitle}{12th AIAA Computational Fluid Dynamics
  Conference}, \bibinfo{year}{1995}{\natexlab{b}}, pp.
  \bibinfo{pages}{806--817}.
\bibitem[{Hirt and Nichols(1981)}]{Hirt1981}
\bibinfo{author}{C.~W. Hirt}, \bibinfo{author}{B.~D. Nichols},
\newblock \bibinfo{title}{{Volume of fluid (VOF) method for the dynamics of
  free boundaries}},
\newblock \bibinfo{journal}{Journal of Computational Physics}
  \bibinfo{volume}{39} (\bibinfo{year}{1981}) \bibinfo{pages}{201--225}.
\bibitem[{Youngs(1982)}]{Youngs1982}
\bibinfo{author}{D.~L. Youngs},
\newblock \bibinfo{title}{{Time-dependent multi-material flow with large fluid
  distortion}},
\newblock in: \bibinfo{booktitle}{Numerical methods for fluid dynamics},
  \bibinfo{publisher}{Academic Press}, \bibinfo{address}{New York},
  \bibinfo{year}{1982}, pp. \bibinfo{pages}{273--486}.
\bibitem[{Lafaurie et~al.(1994)Lafaurie, Nardone, Scardovell, and
  et~al}]{Lafaurie1994}
\bibinfo{author}{B.~Lafaurie}, \bibinfo{author}{C.~Nardone},
  \bibinfo{author}{R.~Scardovell}, \bibinfo{author}{et~al},
\newblock \bibinfo{title}{{ Modeling merging and fragmentation in multiphase
  flows with SURFER}},
\newblock \bibinfo{journal}{International Journal for Numerical Methods in
  Fluids} \bibinfo{volume}{113} (\bibinfo{year}{1994})
  \bibinfo{pages}{134--147}.
\bibitem[{Rudman(1997)}]{Rudman1997}
\bibinfo{author}{M.~Rudman},
\newblock \bibinfo{title}{{Volume-tracking methods for interfacial flow
  calculations}},
\newblock \bibinfo{journal}{International Journal for Numerical Methods in
  Fluids} \bibinfo{volume}{24} (\bibinfo{year}{1997})
  \bibinfo{pages}{671--691}.
\bibitem[{Rider and Kothe(1998)}]{Rider1998}
\bibinfo{author}{W.~J. Rider}, \bibinfo{author}{D.~B. Kothe},
\newblock \bibinfo{title}{{Reconstructing volume tracking}},
\newblock \bibinfo{journal}{Journal of Computational Physics}
  \bibinfo{volume}{141} (\bibinfo{year}{1998}) \bibinfo{pages}{112--152}.
\bibitem[{Pilliod and Puckett(2004)}]{Puckett2004}
\bibinfo{author}{J.~E. Pilliod}, \bibinfo{author}{E.~G. Puckett},
\newblock \bibinfo{title}{{Second-order accurate volume-of-fluid algorithms for
  tracking material interfaces}},
\newblock \bibinfo{journal}{Journal of Computational Physics}
  \bibinfo{volume}{199} (\bibinfo{year}{2004}) \bibinfo{pages}{465--502}.
\bibitem[{Osher and Sethian(1988)}]{Osher1988}
\bibinfo{author}{S.~Osher}, \bibinfo{author}{J.~A. Sethian},
\newblock \bibinfo{title}{{Fronts propagating with curvature-dependent speed:
  Algorithms based on Hamilton-Jacobi formulations}},
\newblock \bibinfo{journal}{Journal of Computational Physics}
  \bibinfo{volume}{79} (\bibinfo{year}{1988}) \bibinfo{pages}{12--49}.
\bibitem[{Osher and Fedkiw(2003)}]{Osher2003}
\bibinfo{author}{S.~Osher}, \bibinfo{author}{R.~Fedkiw}, \bibinfo{title}{{Level
  Set Methods and Dynamic Implicit Surfaces}},
  \bibinfo{publisher}{Springer-Verlag}, \bibinfo{address}{New York},
  \bibinfo{year}{2003}.
\bibitem[{Sussman et~al.(1994)Sussman, Smereka, and Osher}]{Sussman1994}
\bibinfo{author}{M.~Sussman}, \bibinfo{author}{P.~Smereka},
  \bibinfo{author}{S.~Osher},
\newblock \bibinfo{title}{{A Level Set Approach for Computing Solutions to
  Incompressible Two-Phase Flow}},
\newblock \bibinfo{journal}{Journal of Computational Physics}
  \bibinfo{volume}{114} (\bibinfo{year}{1994}) \bibinfo{pages}{146--159}.
\bibitem[{Noh and Woodward(1976)}]{SLIC1976}
\bibinfo{author}{W.~F. Noh}, \bibinfo{author}{P.~Woodward},
\newblock \bibinfo{title}{{SLIC (Simple Line Interface Calculation)}},
\newblock in: \bibinfo{booktitle}{{Fifth International Conference on Numerical
  Methods in Fluid Dynamics}}, \bibinfo{year}{1976}, pp.
  \bibinfo{pages}{330--340}.
\bibitem[{Youngs(1976)}]{Youngs1984}
\bibinfo{author}{D.~L. Youngs}, \bibinfo{title}{{An Interface Tracking Method
  for a 3D Eulerian Hydrodynamics Code}}, \bibinfo{type}{Technical Report}
  \bibinfo{number}{44/92/35}, UK Atomic Weapons Establishment,
  \bibinfo{year}{1976}.
\bibitem[{Puckett(1991)}]{LVIRA}
\bibinfo{author}{E.~G. Puckett},
\newblock \bibinfo{title}{{A volume-of-fluid interface tracking algorithm with
  applications to computing shock wave refraction}},
\newblock in: \bibinfo{booktitle}{Proceedings of the Fourth International
  Symposium on Computational Fluid Dynamics}, \bibinfo{year}{1991}, pp.
  \bibinfo{pages}{933--938}.
\bibitem[{Scardovelli and Zaleski(2003)}]{Scardovelli2003}
\bibinfo{author}{R.~Scardovelli}, \bibinfo{author}{S.~Zaleski},
\newblock \bibinfo{title}{{Interface reconstruction with least-square fit and
  split Eulerian-Lagrangian advection}},
\newblock \bibinfo{journal}{International Journal for Numerical Methods in
  Fluids} \bibinfo{volume}{41} (\bibinfo{year}{2003})
  \bibinfo{pages}{251--274}.
\bibitem[{Aulisa et~al.(2007)Aulisa, Manservisi, Scardovelli, and
  Zaleski}]{Aulisa2007}
\bibinfo{author}{E.~Aulisa}, \bibinfo{author}{S.~Manservisi},
  \bibinfo{author}{R.~Scardovelli}, \bibinfo{author}{S.~Zaleski},
\newblock \bibinfo{title}{{Interface reconstruction with least-squares fit and
  split advection in three-dimensional Cartesian geometry}},
\newblock \bibinfo{journal}{Journal of Computational Physics}
  \bibinfo{volume}{225} (\bibinfo{year}{2007}) \bibinfo{pages}{2301--2319}.
\bibitem[{L\'opez et~al.(2004)L\'opez, Hern\'andez, G\'omez, and
  Faura}]{Lopez2004}
\bibinfo{author}{J.~L\'opez}, \bibinfo{author}{J.~Hern\'andez},
  \bibinfo{author}{P.~G\'omez}, \bibinfo{author}{F.~Faura},
\newblock \bibinfo{title}{{A volume of fluid method based on multidimensional
  advection and spline interface reconstruction}},
\newblock \bibinfo{journal}{Journal of Computational Physics}
  \bibinfo{volume}{195} (\bibinfo{year}{2004}) \bibinfo{pages}{718--742}.
\bibitem[{L\'opez et~al.(2008)L\'opez, Zanzi, G\'omez, and et~al}]{Lopez2008}
\bibinfo{author}{J.~L\'opez}, \bibinfo{author}{C.~Zanzi},
  \bibinfo{author}{P.~G\'omez}, \bibinfo{author}{et~al},
\newblock \bibinfo{title}{{A new volume of fluid method in three
  dimensions-Part \uppercase\expandafter{\romannumeral2}: Piecewise-planar
  interface reconstruction with cubic-B\'ezier fit}},
\newblock \bibinfo{journal}{International Journal for Numerical Methods in
  Fluids} \bibinfo{volume}{58} (\bibinfo{year}{2008})
  \bibinfo{pages}{923--944}.
\bibitem[{Diwakar et~al.(2009)Diwakar, Das, and Sundararajan}]{Diwakar2008}
\bibinfo{author}{S.~V. Diwakar}, \bibinfo{author}{S.~K. Das},
  \bibinfo{author}{T.~Sundararajan},
\newblock \bibinfo{title}{{A Quadratic Spline based Interface (QUASI)
  reconstruction algorithm for accurate tracking of two-phase flows}},
\newblock \bibinfo{journal}{Journal of Computational Physics}
  \bibinfo{volume}{228} (\bibinfo{year}{2009}) \bibinfo{pages}{9107--9130}.
\bibitem[{Xiao et~al.(2005)Xiao, Honma, and Kono}]{THINC}
\bibinfo{author}{F.~Xiao}, \bibinfo{author}{Y.~Honma},
  \bibinfo{author}{T.~Kono},
\newblock \bibinfo{title}{{A simple algebraic interface capturing scheme using
  hyperbolic tangent function}},
\newblock \bibinfo{journal}{International Journal for Numerical Methods in
  Fluids} \bibinfo{volume}{48} (\bibinfo{year}{2005})
  \bibinfo{pages}{1023--1040}.
\bibitem[{Xiao et~al.(2011)Xiao, Ii, and Chen}]{THINCSW}
\bibinfo{author}{F.~Xiao}, \bibinfo{author}{S.~Ii}, \bibinfo{author}{C.~Chen},
\newblock \bibinfo{title}{{Revisit to the THINC scheme: A simple algebraic VOF
  algorithm}},
\newblock \bibinfo{journal}{Journal of Computational Physics}
  \bibinfo{volume}{230} (\bibinfo{year}{2011}) \bibinfo{pages}{7086--7092}.
\bibitem[{Ii et~al.(2012)Ii, Sugiyama, Takeuchi, and et~al}]{MTHINC}
\bibinfo{author}{S.~Ii}, \bibinfo{author}{K.~Sugiyama},
  \bibinfo{author}{S.~Takeuchi}, \bibinfo{author}{et~al},
\newblock \bibinfo{title}{{An interface capturing method with a continuous
  function: The THINC method with multi-dimensional reconstruction}},
\newblock \bibinfo{journal}{Journal of Computational Physics}
  \bibinfo{volume}{231} (\bibinfo{year}{2012}) \bibinfo{pages}{2328--2358}.
\bibitem[{Ii et~al.(2014)Ii, Xie, and Xiao}]{UMTHINC}
\bibinfo{author}{S.~Ii}, \bibinfo{author}{B.~Xie}, \bibinfo{author}{F.~Xiao},
\newblock \bibinfo{title}{{An interface capturing method with a continuous
  function: The THINC method on unstructured triangular and tetrahedral
  meshes}},
\newblock \bibinfo{journal}{Journal of Computational Physics}
  \bibinfo{volume}{259} (\bibinfo{year}{2014}) \bibinfo{pages}{260--269}.
\bibitem[{Xie et~al.(2015)Xie, Ii, and Xiao}]{UMTHINC2}
\bibinfo{author}{B.~Xie}, \bibinfo{author}{S.~Ii}, \bibinfo{author}{F.~Xiao},
\newblock \bibinfo{title}{{An efficient and accurate algebraic interface
  capturing method for unstructured grids in 2 and 3 dimensions: The THINC
  method with quadratic surface representation}},
\newblock \bibinfo{journal}{International Journal for Numerical Methods in
  Fluids} \bibinfo{volume}{76} (\bibinfo{year}{2015})
  \bibinfo{pages}{1025--1042}.
\bibitem[{Xie and Xiao(2017)}]{THINCQQ}
\bibinfo{author}{B.~Xie}, \bibinfo{author}{F.~Xiao},
\newblock \bibinfo{title}{{Toward efficient and accurate interface capturing on
  arbitrary hybrid unstructured grids: The THINC method with quadratic surface
  representation and Gaussian quadrature}},
\newblock \bibinfo{journal}{Journal of Computational Physics}
  \bibinfo{volume}{349} (\bibinfo{year}{2017}) \bibinfo{pages}{415--440}.
\bibitem[{Sethian(1999)}]{FMM}
\bibinfo{author}{J.~A. Sethian}, \bibinfo{title}{{Level Set Methods and Fast
  Marching Methods}}, \bibinfo{publisher}{Cambridge University Press},
  \bibinfo{address}{Cambridge}, \bibinfo{year}{1999}.
\bibitem[{Enright et~al.(2002)Enright, Fedkiw, Ferziger, and
  Mitchell}]{Enright2002}
\bibinfo{author}{D.~Enright}, \bibinfo{author}{R.~Fedkiw},
  \bibinfo{author}{J.~Ferziger}, \bibinfo{author}{I.~Mitchell},
\newblock \bibinfo{title}{{A hybrid particle level set method for improved
  interface capturing}},
\newblock \bibinfo{journal}{Journal of Computational Physics}
  \bibinfo{volume}{183} (\bibinfo{year}{2002}) \bibinfo{pages}{83--116}.
\bibitem[{Sussman and Puckett(2000)}]{Sussman2000}
\bibinfo{author}{M.~Sussman}, \bibinfo{author}{E.~G. Puckett},
\newblock \bibinfo{title}{{A Coupled Level Set and Volume-of-Fluid Method for
  Computing 3D and Axisymmetric Incompressible Two-Phase Flows}},
\newblock \bibinfo{journal}{Journal of Computational Physics}
  \bibinfo{volume}{162} (\bibinfo{year}{2000}) \bibinfo{pages}{301--337}.
\bibitem[{\'Menard et~al.(2007)\'Menard, Tanguy, and Berlemont}]{Menard2007}
\bibinfo{author}{T.~\'Menard}, \bibinfo{author}{S.~Tanguy},
  \bibinfo{author}{A.~Berlemont},
\newblock \bibinfo{title}{{Coupling level set/VOF/ghost fluid methods:
  Validation and application to 3D simulation of the primary break-up of a
  liquid jet}},
\newblock \bibinfo{journal}{International Journal of Multiphase Flow}
  \bibinfo{volume}{33} (\bibinfo{year}{2007}) \bibinfo{pages}{510--524}.
\bibitem[{Yang et~al.(2006)Yang, James, Lowengrub, and et~al}]{ACLSVOF}
\bibinfo{author}{X.~Yang}, \bibinfo{author}{A.~J. James},
  \bibinfo{author}{J.~Lowengrub}, \bibinfo{author}{et~al},
\newblock \bibinfo{title}{{An adaptive coupled level-set/volume-of-fluid
  interface capturing method for unstructured triangular grids}},
\newblock \bibinfo{journal}{Journal of Computational Physics}
  \bibinfo{volume}{217} (\bibinfo{year}{2006}) \bibinfo{pages}{364--394}.
\bibitem[{Sun and Tao(2010)}]{VOSET}
\bibinfo{author}{D.~L. Sun}, \bibinfo{author}{W.~Q. Tao},
\newblock \bibinfo{title}{{A coupled volume-of-fluid and level set (VOSET)
  method for computing incompressible two-phase flows}},
\newblock \bibinfo{journal}{International Journal of Heat \& Mass Transfer}
  \bibinfo{volume}{53} (\bibinfo{year}{2010}) \bibinfo{pages}{645--655}.
\bibitem[{Zhao(2005)}]{FSM}
\bibinfo{author}{H.~K. Zhao},
\newblock \bibinfo{title}{{A fast sweeping method for Eikonal equations}},
\newblock \bibinfo{journal}{Mathematics of Computation} \bibinfo{volume}{74}
  (\bibinfo{year}{2005}) \bibinfo{pages}{603--627}.
\bibitem[{Jiang and Peng(1997)}]{HJWENO}
\bibinfo{author}{G.~S. Jiang}, \bibinfo{author}{D.~Peng},
\newblock \bibinfo{title}{{Weighted ENO Schemes for Hamilton-Jacobi
  Equations}},
\newblock \bibinfo{journal}{SIAM Journal on Scientific Computing}
  \bibinfo{volume}{21} (\bibinfo{year}{1997}) \bibinfo{pages}{2126--2143}.
\bibitem[{Shu and Osher(1989)}]{TVDRK3}
\bibinfo{author}{C.~W. Shu}, \bibinfo{author}{S.~Osher},
\newblock \bibinfo{title}{{Efficient implementation of essentially
  non-oscillatory shock-capturing schemes}},
\newblock \bibinfo{journal}{Journal of Computational Physics}
  \bibinfo{volume}{83} (\bibinfo{year}{1989}) \bibinfo{pages}{32--78}.
\bibitem[{Gottlieb(2005)}]{SSPRK}
\bibinfo{author}{S.~Gottlieb},
\newblock \bibinfo{title}{{On high order strong stability preserving
  runge-kutta and multi step time discretizations}},
\newblock \bibinfo{journal}{SIAM Journal on Scientific Computing}
  \bibinfo{volume}{25} (\bibinfo{year}{2005}) \bibinfo{pages}{105--128}.
\bibitem[{Harvie and Fletcher(2000)}]{Harvie2000}
\bibinfo{author}{D.~J.~E. Harvie}, \bibinfo{author}{D.~F. Fletcher},
\newblock \bibinfo{title}{{A New Volume of Fluid Advection Algorithm: The
  Stream Scheme}},
\newblock \bibinfo{journal}{Journal of Computational Physics}
  \bibinfo{volume}{162} (\bibinfo{year}{2000}) \bibinfo{pages}{1--32}.
\bibitem[{Zalesak(1979)}]{Zalesak}
\bibinfo{author}{S.~T. Zalesak},
\newblock \bibinfo{title}{{Fully multidimensional flux-corrected transport
  algorithms for fluids}},
\newblock \bibinfo{journal}{Journal of Computational Physics}
  \bibinfo{volume}{31} (\bibinfo{year}{1979}) \bibinfo{pages}{335--362}.
\bibitem[{Harvie and Fletcher(2001)}]{Harvie2001}
\bibinfo{author}{D.~J.~E. Harvie}, \bibinfo{author}{D.~F. Fletcher},
\newblock \bibinfo{title}{{A new volume of fluid advection algorithm: the
  defined donating region scheme}},
\newblock \bibinfo{journal}{International Journal for Numerical Methods in
  Fluids} \bibinfo{volume}{35} (\bibinfo{year}{2001})
  \bibinfo{pages}{151--172}.
\bibitem[{Yokoi(2007)}]{THINCWLIC}
\bibinfo{author}{K.~Yokoi},
\newblock \bibinfo{title}{{Efficient implementation of THINC scheme: A simple
  and practical smoothed VOF algorithm}},
\newblock \bibinfo{journal}{Journal of Computational Physics}
  \bibinfo{volume}{226} (\bibinfo{year}{2007}) \bibinfo{pages}{1985--2002}.
\bibitem[{L\'opez et~al.(2005)L\'opez, Hern\'andez, G\'omez, and
  Faura}]{Lopez2005}
\bibinfo{author}{J.~L\'opez}, \bibinfo{author}{J.~Hern\'andez},
  \bibinfo{author}{P.~G\'omez}, \bibinfo{author}{F.~Faura},
\newblock \bibinfo{title}{{An improved PLIC-VOF method for tracking thin fluid
  structures in incompressible two-phase flows}},
\newblock \bibinfo{journal}{Journal of Computational Physics}
  \bibinfo{volume}{208} (\bibinfo{year}{2005}) \bibinfo{pages}{51--74}.
\bibitem[{Aulisa et~al.(2003)Aulisa, Manservisi, and Scardovelli}]{Aulisa2003}
\bibinfo{author}{E.~Aulisa}, \bibinfo{author}{S.~Manservisi},
  \bibinfo{author}{R.~Scardovelli},
\newblock \bibinfo{title}{{A mixed markers and volume-of-fluid method for the
  reconstruction and advection of interfaces in two-phase and free-boundary
  flows}},
\newblock \bibinfo{journal}{Journal of Computational Physics}
  \bibinfo{volume}{188} (\bibinfo{year}{2003}) \bibinfo{pages}{611--639}.

\end{thebibliography}

\end{document}